\documentclass[aps,pra,amsmath,amssymb,amsfonts,lengthcheck,twocolumn,superscriptaddress]{revtex4-2}
\usepackage{graphicx}
\usepackage{subfigure}
\usepackage{amsthm}
\usepackage{verbatim}
\usepackage{dcolumn}
\usepackage{bm}
\usepackage{epsf}
\usepackage{color}
\usepackage[colorlinks=true,citecolor=blue,linkcolor=blue,urlcolor=blue]{hyperref}%
\usepackage{xcolor}
\usepackage{dsfont}
\usepackage{tikz}
\usepackage{todonotes}
\usepackage{multirow}
\usepackage{xfrac}
\usepackage{makecell}

\newcommand{\bra}[1]{\left\langle #1\right|}
\newcommand{\ket}[1]{\left|#1\right\rangle}
\newcommand{\braket}[2]{\left\langle #1|#2\right\rangle}

\newcommand{\ptr}[2]{\mathrm{tr_{#1}}\left\{#2\right\}}

\newcommand{\la}{\left\langle}
\newcommand{\ra}{\right\rangle}
\newcommand{\pd}{\partial}

\newcommand{\e}[1]{\exp{\left(#1\right)}}

\newcommand{\id}{\mathbb{I}}
\newcommand{\com}[2]{\left[#1,\,#2\right]}

\newcommand{\co}[1]{\cos{\left(#1\right)}}
\newcommand{\si}[1]{\sin{\left(#1\right)}}

\newcommand{\ch}[1]{\cosh{\left(#1\right)}}

\newcommand{\tah}[1]{\tanh{\left(#1\right)}}
\newcommand{\bla}{bla\\bla\\bla\\bla\\bla}

\newcommand{\mc}[1]{\mathcal{#1}}

\newcommand{\mf}[1]{\mathfrak{#1}}
\newcommand{\mrm}[1]{\mathrm{#1}}

\DeclareMathAlphabet\mathbfcal{OMS}{cmsy}{b}{n}

\makeatletter
\newcommand{\currentfontsize}{\f@size pt}
\makeatother

\makeatletter
\newcommand\footnoteref[1]{\protected@xdef\@thefnmark{\ref{#1}}\@footnotemark}
\makeatother

\begin{document}

\title{Enhancing precision thermometry with nonlinear qubits}

\author{Sebastian Deffner}
\email{deffner@umbc.edu}
\affiliation{Department of Physics, University of Maryland, Baltimore County, Baltimore, MD 21250, USA}
\affiliation{National Quantum Laboratory, College Park, MD 20740, USA}

\begin{abstract}
Quantum thermometry refers to the study of measuring ultra-low temperatures in quantum systems. The precision of such a quantum thermometer is limited by the degree to which temperature can be estimated by quantum measurements. More precisely, the maximal precision is given by the inverse of the quantum Fisher information. In the present analysis, we show that quantum thermometers that are described by nonlinear Schr\"odinger equations allow for a significantly enhanced precision, that means larger quantum Fisher information. This is demonstrated for a variety of pedagogical scenarios consisting of single and two-qubits systems. The enhancement in precision is indicated by non-vanishing quantum speed limits, which originate in the fact that the thermal, Gibbs state is typically not invariant under the nonlinear equations of motion.
\end{abstract}

\maketitle

\section{Introduction}

The year 2024 marks the second centennial of the birth of thermodynamics, which is commonly associated with the publication of Carnot's ``Reflections on the Motive Power of Fire'' \cite{Carnot1824}. The ``Carnot efficiency'' quantifies the maximal amount of work any (cyclic) thermodynamic process can extract from the heat flow between two reservoirs of different temperatures. Namely, the extractable work is at least $T_\mrm{cold}/T_\mrm{cold}$ smaller than the exchanged heat, where $T_\mrm{hot}$ and $T_\mrm{cold}$ are hot and cold temperatures, respectively. Operationally, this means that temperature ratios can be measured by recording the work output of Carnot engines \cite{Callen1985}. 

Nevertheless, the invention of thermometers predates thermodynamics by at least two centuries, of which Galileo's thermoscope is arguably the most important milestone \cite{fretwell1937development}. The first modern thermometer was invented by Fahrenheit, who quickly realized that the mercury-in glass version gave much higher precision than working with wine spirit \cite{fretwell1937development}. Nowadays, mercury thermometers have fallen out of fashion due to the toxicity of mercury to humans and the environment. Interestingly, mercury was also the first element, for which superconductivity was discovered \cite{Tresca2022PRB}. In the low temperature limit, superconductivity can described by the Ginzburg-Landau theory \cite{Ashcroft1976}, which in mean-field gives rise to nonlinear Schr\"odinger equations \cite{huang2009introduction}. The natural question arises, how the precision of thermometers at such low, superconducting temperatures is affected by the effectively nonlinear dynamics.

This question becomes particularly interesting in context of our recent result showing that nonlinear dynamics allow for faster evolution of quantum states \cite{Deffner2022EPL}. The maximal rate of quantum evolution is determined by the quantum speed limit (QSL), which is a more careful formulation of the Heisenberg uncertainty relation of energy and time \cite{Poggi2013EPL,Poggi2016PRA,Deffner2017JPA}. In Ref.~\cite{Deffner2022EPL} we showed that the QSL grows as a function of the ``strength'' of the nonlinearity, yet also that the QSL depends non-trivially on the polynomial power of the nonlinear term. Since it also has been elucidated that the QSL is intimately related to quantum metrology \cite{Giovannetti2011NPho}, it is not far-fetched to realize that one should be able to leverage nonlinear quantum evolution as a resource in quantum thermometry. For instance, related conclusions were drawn about parameter estimation in complex quantum many body system \cite{Boixo2007PRL,Roy2008PRL,Beau2017PRL}, such as the long-range Kitaev chains \cite{Yang2022PRR}.

However, addressing this question is far from trivial as the temperature of quantum systems is neither a classical nor a quantum observable, but rather a parameter that has to be estimated \cite{Campbell2018QST,Mehboudi2019JPA}. In fact, quantum thermometry has attracted significant research efforts in the quest to identify \emph{optimal} measurement schemes \cite{Campbell2017NJP,Kiilerich2018PRA,Seveso2018PRA,Razavian2019,Mukherjee2019CP,Feyles2019PRA,Mitchison2020PRL,Connor2012Entropy,HovhannisyanPRXQ2021,Mok2021CP,Sekatski2022optimal,
Albarelli2023PRA,Mihailescu2023PRA,Mihailescu2024QST} and genuinely quantum resources \cite{Brunelli2011PRA,Brunelli2012PRA,Cavina2018PRA,Potts2019fundamentallimits,Genoni2019JPA,Montenegro2020PRR,Mancino2020PRR,Candeloro2021PRE,
Salado-Mejía2021QST,Brenes2023PRA,Sone2024QST,Aiache2024PRE} that could be leveraged to build thermometers with improved precision at ultralow temperatures.

In the present work, we show that nonlinear quantum dynamics can, indeed, be leveraged to increase the precision of thermometry. More specifically, we demonstrate that the quantum Fisher information \cite{Liu2020JPA} for temperature, can be enhanced in nonlinear quantum systems. To this end, we analyze single and two-qubit systems undergoing arbitrary nonlinear dynamics. 

As a first result, we corroborate our findings of Ref.~\cite{Deffner2022EPL}, namely we show that in isolated systems the QSL grows as a function of the strength of the nonlinear term in the evolution equation. More importantly, we also show that for situations for which the thermal state is not stationary under the nonlinear dynamics, the quantum Fisher information is a growing function of time. In other words, we show that thermometers with nonlinear working mediums allow for higher precision. This effect is even more pronounced in our two-qubit example, which is a more realistic description of thermometry. In this case, we find that while the QSL is reduced, the quantum Fisher information dramatically grows as a function of time. This behavior can be traced back to the effective ``twisting'' of the quantum state around the Bloch sphere, which is driven by the nonlinear term in the dynamics.

In very simple terms, our results suggest that mercury-in glass thermometers may have an even higher precision below the transition to superconductity. Curiously, appropriately modernized versions of the first thermometer might, hence, still give the highest precision.

\section{Preliminaries}

We start by briefly reviewing the main notions and concepts, and by establishing our notation.

\paragraph*{Quantum speed limit}

As mentioned above, we recently showed \cite{Deffner2022EPL} that nonlinear quantum dynamics can be exploited to enhance the speed of quantum state evolution. The maximal speed is given by the QSL, which was originally derived for simple, undriven Schr\"odinger dynamic by Mandelstam and Tamm \cite{Mandelstam1945}. They showed that the minimal time a quantum system needs to evolve between orthogonal states is bounded from below by the variance of the energy, $\Delta E$, and hence $\tau_\mrm{QSL}=\pi\hbar/2\Delta E$. Since, however, the variance of an operator is not necessarily a good quantifier for undriven dynamics \cite{Uffink1993}, Margolus and Levitin \cite{margolus98} revisited the problem and derived a second bound on the quantum evolution time in terms of the average energy $E=\la H\ra-E_g$ over the ground state with energy $E_g$, that means $\tau_\mrm{QSL}=\pi\hbar/2E$. Interestingly, the combined bound can be shown to be tight and attainable \cite{Levitin2009}.

More recently, it has been recognized that the QSLs are bounds on the rate with which quantum states become distinguishable \cite{Fogarty2020PRL,Poggi2021PRQ,Campo2021PRL}. Therefore, one typically considers different geometric measures of distinguishability \cite{Pires2016PRX,OConnor2021PRA} in the derivation of QSL. In our work, we have shown that these different treatments become equivalent when considering the metric properties of the quantum dynamics \cite{Deffner2017NJP,Deffner2020PRR,Aifer2022NJP}.

In the present analysis, we work with the simplest version of the QSL \cite{Deffner2017NJP}, namely
\begin{equation}
\label{eq:vQSL}
v_\mrm{QSL}\equiv ||\dot{\rho}_t||\,,
\end{equation}
where $\rho_t$ is a time-dependent quantum state, and $||\cdot ||$ is the operator norm. For linear, unitary dynamics we simply have $v_\mrm{QSL}=\la H^2_t\ra/\hbar^2$. where $H_t$ is the possibly time-dependent Hamiltonian of the system of interest. In Ref.~\cite{Deffner2022EPL} we then showed that $v_\mrm{QSL}$ increases monotonically with the strength of the nonlinearity for harmonic oscillators and time-dependent boxes under going Gross-Pitaeveskii \cite{Gross1961,Pitaevskii1961} and Kolomeisky \cite{Kolomeisky2000PRL} dynamics.

Here, we focus on qubits systems, for which analytical solutions for many cases can be found, cf. Refs.~\cite{Barnes2012PRL,Barnes2013PRA}. The quantum state of a single qubit can be written in its Bloch representation as
\begin{equation}
\rho_t=\frac{1}{2}\left(\id_2+\vec{r}_t \cdot \vec{\sigma}\right)\,,
\end{equation}
where $\vec{\sigma}=(\sigma_x,\sigma_y,\sigma_z)$ is the Pauli vector, and $\vec{r}_t=(x_t,y_t,z_t)$ denotes the Bloch vector. In this representation, the QSL \eqref{eq:vQSL} simply becomes,
\begin{equation}
\label{eq:qsl_qubit}
v_\mrm{QSL}=\frac{1}{2} \sqrt{\left|\dot{x}_t^2+\dot{y}_t^2+\dot{z}_t^2\right|}\,.
\end{equation}
As we will see below, this expression \eqref{eq:qsl_qubit} makes the influence of nonlinear terms in the dynamics on the QSL particularly transparent.

\paragraph*{Quantum metrology -- parameter estimation}

While the QSL for nonlinear dynamics is interesting, the focus of the present analysis is on quantum thermometry. In general quantum settings, the temperature is a parameter that needs to be estimated based on measurement outcomes. The maximal precision of such an estimation is given by the Cramer-Rao bound \cite{Liu2020JPA}. For generality, we consider a quantum state, $\rho(\vec{\lambda})$, which encodes a set of $n$ parameters, $\vec{\lambda}=(\lambda_1,\cdots,\lambda_n)$. The precision with which these parameters can be estimated is quantified by the elements of the covariance matrix $\mathbf{Cov}[\vec{\lambda}]$, which is defined by
\begin{equation}
\text{Cov}(\lambda_i,\lambda_j)\equiv \la(\lambda_i-\la\lambda_i\ra) (\lambda_j-\la\lambda_j\ra)\ra\,.
\end{equation}
Note that the diagonal elements of the covariance matrix are simply the variances of the single parameters, $\lambda_i$.

The multiparameter Cramer-Rao bound \cite{Liu2020JPA} states that the covariance matrix is lower bounded by the inverse of the quantum Fisher information matrix, $\mathbfcal{F}$, \cite{Liu2020JPA}
\begin{equation}
\label{eq:CR}
\mathbf{Cov}[\vec{\lambda}]\geq \frac{1}{N} \mathbfcal{F}^{-1}\,,
\end{equation}
where $N$ is the number of measurements taken on the quantum state, $\rho(\vec{\lambda})$. It has been noted that Eq.~\eqref{eq:CR} only gives meaningful insight if $\mathbfcal{F}$ is invertible. For instance, if there are only two parameters, which are not independent from each other, $\mathbfcal{F}$ may become singular \cite{Mihailescu2024QST}. In such cases, Eq.~\eqref{eq:CR} has to be considered component-by-component, noting that the single-parameter Cramer-Rao bound can not necessarily be interpreted as setting the ultimate limit on the precision of measurements \cite{mihailescu2024uncertain}.

For general quantum states, the quantum Fisher information matrix is defined as
\begin{equation}
\label{eq:fisher}
\mc{F}_{\nu,\mu}= 2\sum_{k,l} \frac{\mathfrak{R}\left(\bra{l}\pd_\nu \rho\ket{k}\bra{k}\pd_\mu \rho\ket{l}\right)}{p_l+p_k}\,,
\end{equation}
where $\{p_k, \ket{k}\}$ is the eigensystem of $\rho$. Moreover, $\pd_\nu$ denotes the partial derivative with respect to the $\nu$th parameter. The expression for  $\mc{F}_{\nu,\mu}$ \eqref{eq:fisher} becomes much simpler for qubit states, and we have  \cite{Liu2020JPA}
\begin{equation}
\label{eq:fisher_qubit}
\mc{F}_{\nu,\mu}=(\pd_\nu \vec{r})\cdot (\pd_\mu \vec{r}_t) +\frac{(\vec{r}\cdot \pd_\nu \vec{r})\,(\vec{r}\cdot \pd_\mu \vec{r})}{1-|\vec{r}|^2}\,,
\end{equation}
where $\vec{r}$ is again the Bloch vector.

\paragraph*{Nonlinear quantum dynamics}

To finally study quantum thermometry in nonlinear systems, we need to define the corresponding dynamics. To this end, we consider general quantum dynamics described by the nonlinear Schr\"odinger equation \cite{Childs2016PRA}
\begin{equation}
i\hbar\, \pd_t \ket{\psi_t}=H_t\,\ket{\psi_t}+ K \ket{\psi_t}
\end{equation}
where $H_t$ is the usual, Hermitian, possibly time-dependent Hamiltonian. Further, $K$ describes a nonlinearity of the form
\begin{equation}
\bra{x} \left(K \ket{\psi_t}\right)= \mc{K}(|\braket{x}{\psi_t}|)\,\braket{x}{\psi_t}\,.
\end{equation}
For instance, for Gross-Piateveskii dynamics \cite{Gross1961,Pitaevskii1961} we have $\mc{K}(|\braket{x}{\psi_t}|)=\kappa\,|\braket{x}{\psi_t}|^2$, and for Kolomeisky dynamics \cite{Kolomeisky2000PRL} $\mc{K}(|\braket{x}{\psi_t}|)=\kappa\,|\braket{x}{\psi_t}|^4$. However, also more complicated nonlinearities have been considered, as for instance logarithmic terms \cite{Childs2016PRA}. The Gross-Pitaevskii equation \cite{Gross1961,Pitaevskii1961} is probably the most widely know nonlinear Schr\"odinger equation, with widespread applications from Bose-Einstein condensation \cite{Gross1961,Pitaevskii1961} over nonlinear optics \cite{Rand2010} to  plasma physics \cite{Ruderman2002}. Logarithmic nonlinearities were discussed in the context of Bose liquids \cite{Meyer2014PRA}. 

Again considering qubits, the nonlinear term, $\mc{K}_t$, drastically simplifies. Since $|\braket{0}{\psi_t}|^2=(1+z_t)/2$ and $|\braket{1}{\psi_t}|^2=(1-z_t)/2$, $\mc{K}_t$ can be written as an effectively state-dependent Hamiltonian \cite{Childs2016PRA},
\begin{equation}
\mc{K}_t=\begin{pmatrix}
\kappa\left(\sqrt{(1+z_t)/2}\right)&0\\
0&\kappa\left(\sqrt{(1-z_t)/2}\right)\,.
\end{pmatrix}
\end{equation}
Note that $\mc{K}_t$ is diagonal in the Bloch representation for all choices of the Hamiltonian $H_t$. Therefore, we immediately observe that states that are diagonal in the Bloch representation with constant $z_t$ are not affected, i.e., invariant under the nonlinear term, $\mc{K}_t$.

In the following, we will now analyze the quantum speed limit \eqref{eq:qsl_qubit} as well as the quantum Fisher information matrix \eqref{eq:fisher_qubit} for several, relevant choices of the Hamiltonian $H_t$. To allow for thermometry, we will further assume that the qubit is initially prepared in a thermal, Gibbs state.

\section{Deliberations for single qubits}

We start with analyzing the dynamics and thermometric properties of single, nonlinear qubits. For the sake of clarity and simplicity we further assume that the self-Hamiltonian of the qubits is time-independent.

\subsection{Spin-1/2 particle in magnetic field}

Arguably, the simplest scenario is a spin-1/2 particle in a one-dimensional magnetic field. The corresponding Hamiltonian reads,
\begin{equation}
H=\Delta\, \sigma_z\,,
\end{equation}
where $\Delta \geq 0$. It is then a simple exercise to show that the von-Neumann equation
\begin{equation}
\label{eq:vonNeumann}
i\hbar\, \pd_t\rho_t=\com{H+\mc{K}_t}{\rho_t}
\end{equation} 
is equivalent to 
\begin{equation}
\label{eq:diffeq}
\left(\dot{x}_t,\dot{y}_t,\dot{z}_t\right)=\left[2\Delta +\tilde{\kappa}(z_t)\right] \,\left(y_t,-x_t,0\right)
\end{equation}
where $\tilde{\kappa}(z)=\kappa(\sqrt{(1+z)/2})-\kappa(\sqrt{(1-z)/2})$. Note that for $\Delta=0$ the differential Eqs.~\eqref{eq:diffeq} are identical to the dynamics analyzed in Ref.~\cite{Childs2016PRA}. 

Equations~\eqref{eq:diffeq} can be solved analytically, and we have
\begin{equation}
\label{eq:solution_simple}
\begin{aligned}[b]
x_t&= x_0 \co{\left[2 \Delta + \tilde{\kappa}(z_0)\right] t} +y_0 \si{\left[2 \Delta + \tilde{\kappa}(z_0)\right] t} \\
y_t&=y_0 \co{\left[2 \Delta + \tilde{\kappa}(z_0)\right] t} -x_0 \si{\left[2 \Delta + \tilde{\kappa}(z_0)\right] t}\\
z_t&= z_0\,,
\end{aligned}
\end{equation}
which hold true for any choice of the nonlinearity $\mc{K}_t$. From these solutions \eqref{eq:solution_simple}, we can directly compute an expression for the QSL \eqref{eq:qsl_qubit}. We have,
\begin{equation}
v_\mrm{QSL}=\frac{1}{2} \sqrt{x_0^2+y_0^2}\,\left|2 \Delta +\tilde{\kappa}(z_0)\right|\,,
\end{equation}
which corroborates our earlier findings in Ref.~\cite{Deffner2022EPL}. Namely, the QSL grows as a function of the magnitude of the nonlinearity $\tilde{\kappa}(z_0)$. However, we also immediately recognize that the QSL is different from zero only for such initial states for which $x_0\neq 0$ or $y_0\neq 0$.

For this simple qubit, the thermal Gibbs state, $\rho_\mrm{Gibbs}\propto \e{-\beta H}$, is diagonal in the Bloch representation, and we have thus have $\vec{r}_t=(0,0,z_0)$. In other words, the Gibbs state remains stationary also under the nonlinear dynamics \eqref{eq:vonNeumann}. Moreover, the quantum Fisher information matrix \eqref{eq:fisher_qubit} simplifies to read
\begin{equation}
\mc{F}_{\nu,\mu}=\frac{\pd_\nu z_0\, \pd_\mu z_0}{1-z_0^2}\,,
\end{equation}
where $\nu,\mu \in \{\beta,\Delta\}$. In principle, the energy splitting $\Delta$ and the inverse temperature $\beta$ are independent parameters. However, the thermal state only depends on the product $\beta\Delta$, and not on the parameters separately, cf. Ref.~\cite{Campbell2018QST}. We conclude that the quantum Fisher information matrix is (i) singular as $\mrm{det}(\mc{F})=0$, and (ii) constant under the nonlinear dynamics \eqref{eq:vonNeumann}. Thus, there is no advantage in the precision of thermometry originating in the nonlinear term. As we will see shortly, this is a peculiarity of this simplest qubit.

\subsection{Landau-Zener model}

The situation becomes more interesting for the paradigmatic Landau-Zener model \cite{landau1932theorie,zener1932non,stuckelberg1932theorie} with Hamiltonian,
\begin{equation}
\label{eq:H_LZ}
H=\Delta\, \sigma_z+ J \sigma_x\,.
\end{equation}
In this case, the dynamics is described by
\begin{equation}
\label{eq:diffeq_LZ}
\begin{aligned}[b]
\dot{x}_t&=\quad \left[2\Delta +\tilde{\kappa}(z_t)\right]\,y_t\\
\dot{y}_t&=\,-\,\left[2\Delta +\tilde{\kappa}(z_t)\right]x_t+2 J z_t\\
\dot{z}_t&=\,-\, 2 J y_t\,.
\end{aligned}
\end{equation}
Already this slightly more complicated situation is no longer analytically solvable, however it remains easily tractable numerically. To this end, we now continue with a Landau-Zener qubit that is initially prepared in a Gibbs state, which can be written in the Bloch representation, $\rho_\mrm{Gibbs}\equiv\left(\id_2+\vec{r}_\mrm{Gibbs}\cdot\vec{\sigma}\right)/2$, as
\begin{equation}
\label{eq:Gibbs_LZ}
\vec{r}_\mrm{Gibbs}=-(J,0,\Delta)\,\frac{\tanh{\left(\beta\,\sqrt{J^2+\Delta^2}\right)}}{\sqrt{J^2+\Delta^2}}\,.
\end{equation}
Note that this quantum state is not diagonal in the Bloch representation, and hence the Gibbs state can no longer be stationary under the nonlinear dynamics \eqref{eq:vonNeumann}.

\begin{figure*}
\includegraphics[width=.48\textwidth]{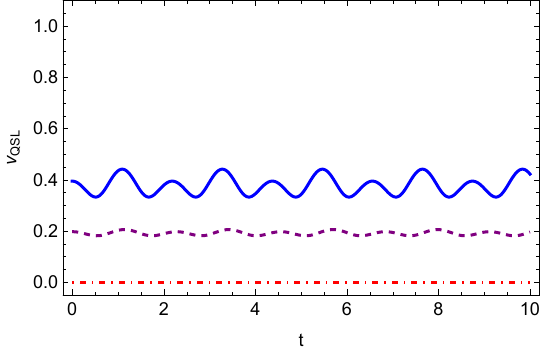}\hfill \includegraphics[width=.48\textwidth]{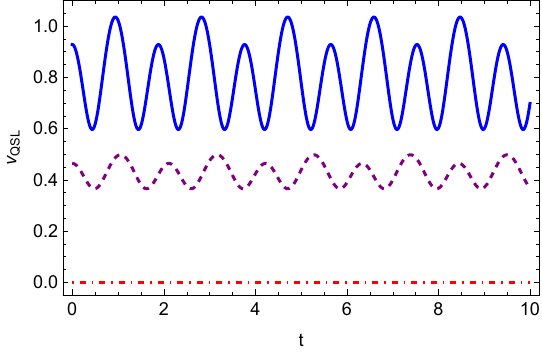}
\caption{\label{fig:qsl_LZ} Quantum speed limit \eqref{eq:qsl_qubit} for the nonlinear Landau-Zener model \eqref{eq:diffeq_LZ} with a Gross-Pitaevskii nonlinearity, $\kappa(x)=g x^2$, (left panel) and a logarithmic nonlinearity, $\kappa(x)=g \ln(x^2)$, (right panel) with an initial Gibbs state \eqref{eq:Gibbs_LZ}. Parameters are $\Delta=1$, $J=1$, $\beta=1$, for $g= 0$ (red, dotdashed line), $g=1$ (purple, dashed line), and $g= 2$ (blue, solid line).}
\end{figure*}

In Fig.~\ref{fig:qsl_LZ} we depict the resulting QSL \eqref{eq:qsl_qubit} for a Gross-Pitaevskii nonlinearity, $\kappa(x)=g x^2$, as well as a logarithmic term, $\kappa(x)=g \ln(x^2)$. As expected, $v_\mrm{QSL}>0$ for nonlinear dynamics, since the Gibbs state is no longer stationary. However, we also observe that once again (i) nonlinear terms enhance the rate of quantum state evolution, and (ii) the nonlinear speed-up is more pronounced for the logarithmic case. This is somewhat expected, as for small arguments the magnitude of logarithm $\log{x^2}$ diverges.

\paragraph*{Thermometry on nonlinear Landau-Zener qubits}

When considering the Cramer-Rao bound \eqref{eq:CR} for the Landau-Zener model, the first observation is that there are now three parameters, $\Delta$, $J$, and $\beta$. The corresponding quantum Fisher information matrix \eqref{eq:fisher_qubit} is no longer singular, which can be directly read off from its determinant,
\begin{widetext}
\begin{equation}
\begin{split}
&\mrm{det}(\mathbfcal{F})=\frac{J^2 (\beta +\Delta ) }{2 \left(\Delta
   ^2+J^2\right)^3} \tanh ^2\left(\beta 
   \sqrt{\Delta ^2+J^2}\right) \text{sech}^4\left(\beta  \sqrt{\Delta ^2+J^2}\right)\\
   & \times \left[-\Delta +2 \beta  \left(\Delta ^2+J^2\right) \left(\Delta  (\Delta -\beta
   )+J^2\right)+\Delta  \cosh \left(2 \beta  \sqrt{\Delta ^2+J^2}\right)\right]\,.
\end{split}
\end{equation}
\end{widetext}

\begin{figure*}
\includegraphics[width=.48\textwidth]{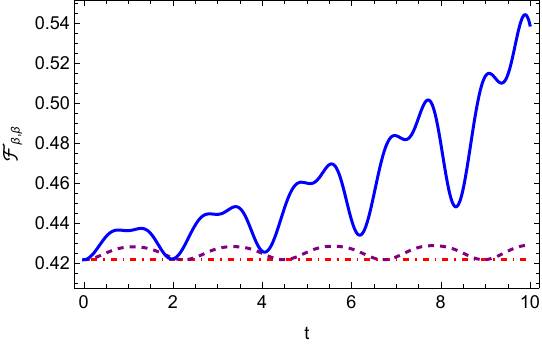}\hfill \includegraphics[width=.48\textwidth]{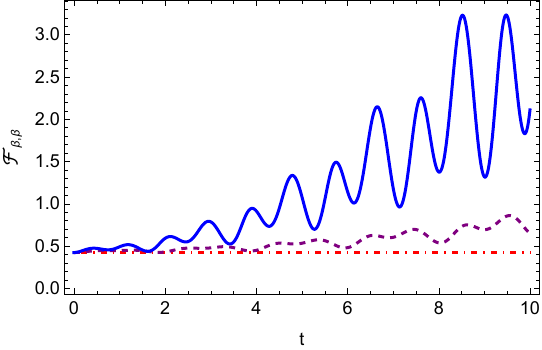}
\caption{\label{fig:fish_LZ} Quantum Fisher information for temperature, $\mc{F}_{\beta,\beta}$, \eqref{eq:fisher_qubit} for the nonlinear Landau-Zener model \eqref{eq:diffeq_LZ} with a Gross-Pitaevskii nonlinearity, $\kappa(x)=g x^2$, (left panel) and a logarithmic nonlinearity, $\kappa(x)=g \ln(x^2)$, (right panel) with an initial Gibbs state \eqref{eq:Gibbs_LZ}. Parameters are $\Delta=1$, $J=1$, $\beta=1$, for $g= 0$ (red, dotdashed line), $g=1$ (purple, dashed line), and $g= 2$ (blue, solid line).}
\end{figure*}

Second, since the Gibbs state is not stationary under the nonlinear dynamics, also $\mathbfcal{F}$ becomes a time-dependent function. Solving the evolution equations \eqref{eq:diffeq_LZ} numerically for a range of values of the inverse temperature $\beta$, we computed the matrix element $\mc{F}_{\beta,\beta}$. Our results are shown in Fig.~\ref{fig:fish_LZ} again for the Gross-Pitaevskii as well as the logarithmic nonlinearity. We observe that  for every instant $t$ the quantum Fisher information $\mc{F}_{\beta,\beta}$ grows as a function of the strength of the nonlinear term. Moreover, we see that as a function of time the amplitude of $\mc{F}_{\beta,\beta}$ grows, and that the effect is more pronounced for the logarithmic case. Hence, we conclude that nonlinear quantum effects are, indeed, a resource that can be leveraged to enhance the precision of thermometry. Note, however, that for the Landau-Zener model we have studied a genuinely non-equilibrium situations, as the Gibbs state is not invariant under the nonlinear dynamics \eqref{eq:vonNeumann}. Therefore, we now continue with a more realistic description of a quantum thermometer with nonlinear working medium.

\section{Nonlinear quantum thermometry}

Considering a single qubit in isolation, yet prepared in a thermal, Gibbs state is only a poor representation of actually measuring the temperature of an object. More realistically, a quantum system has equilibrated with a thermal environment, and the temperature is measured by letting the system interact with a thermometer. The simplest scenario is then a ``system'' qubit, $\mf{S}$, that is in interaction with a ``thermometer'' qubit, $\mf{T}$. 

In Ref.~\cite{Campbell2018QST} we considered exactly this situation for linear dynamics. For our present purposes, we now assume that the thermometer is a nonlinear qubit. The total Hamiltonian then reads
\begin{equation}
\label{eq:thermo}
H=\left(H^\mf{T}+\mc{K}_t^\mf{T}\right)\otimes \id^\mf{S}+\id^\mf{T}\otimes H^\mf{S}+h_\mrm{int}\,,
\end{equation}
where the bare Hamiltonians are taken to be spin-1/2 particles
\begin{equation}
H^\mf{S}=H^\mf{T}=\Delta\, \sigma_z\,.
\end{equation}
In complete analogy to Ref.~\cite{Campbell2018QST} we choose the interaction to be
\begin{equation}
\label{eq:interaction}
h_\mrm{int}=\frac{J}{2}\left(\sigma_x\otimes \sigma_x+\sigma_y\otimes\sigma_y \right)\,,
\end{equation}
which in the linear case describes a state-swap.

Despite its simplicity, such a scenario is not totally unrealistic. For instance, one could imagine the nonlinear thermometer to be built from a Bose-Einstein condensate in an external double well potential \cite{Byrnes2015}. The interaction term \eqref{eq:interaction} can be easily facilitated with standard quantum logic gates.

A general, time-dependent two-qubit quantum state can be written as
\begin{equation}
\label{eq:2q}
\rho_t=\frac{1}{4} \sum_{i,j\in \{0,x,y,z\}} \alpha_{i,j}(t)\, \sigma_i\otimes \sigma_j\,,
\end{equation}
which is a generalized Bloch representation, and $\sigma_0\equiv \id_2$. For quantum thermometry, we now assume that the system qubit $\mf{S}$ to be prepared its the thermal Gibbs state, and $\mf{T}$ is initially in its ground state,
\begin{equation}
\label{eq:rho0}
\rho_0=\rho_0^\mf{T}\otimes\rho_0^\mf{S}=\ket{0}\bra{0}\otimes \frac{\e{-\beta \Delta \sigma_z}}{2 \ch{\beta \Delta}}\,.
\end{equation}
As above, we now have to solve for the dynamics of the joint system. For the sixteen, time-dependent coefficients, $\alpha_{i,j}(t)$,  the evolution equations can be written in reasonably clean form, which we have collected in the Appendix \ref{sec:appA}. Remarkably, the dynamics is analytically solvable for the linear case, but for the nonlinear thermometer we again have to resort to numerics.

We are now interested in the dynamics of the thermometer $\mf{T}$, and with what precision the temperature of $\mf{S}$ can be read off. To this end, we need to compute the QSL \eqref{eq:vQSL} as well as the quantum Fisher information \eqref{eq:fisher_qubit} from the reduced state of the thermometer,
\begin{equation}
\label{eq:T}
\rho_t^\mf{T}=\ptr{\mf{S}}{\rho_t}=\frac{1}{2}\left(\id_2 +\vec{r}_t^\mf{T} \cdot \vec{\sigma}\right)\,,
\end{equation}
where $\vec{r}_t^\mf{T}=(\alpha_{x,0}(t),\alpha_{y,0}(t),\alpha_{z,0}(t))$.

\begin{figure*}
\includegraphics[width=.48\textwidth]{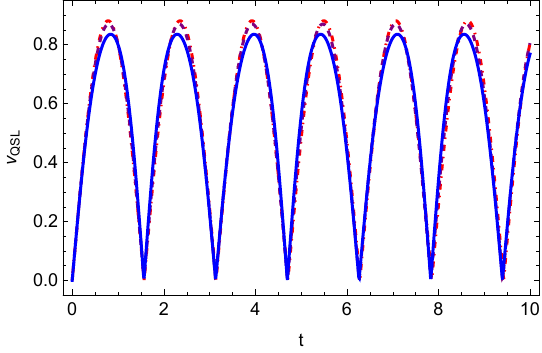}\hfill \includegraphics[width=.48\textwidth]{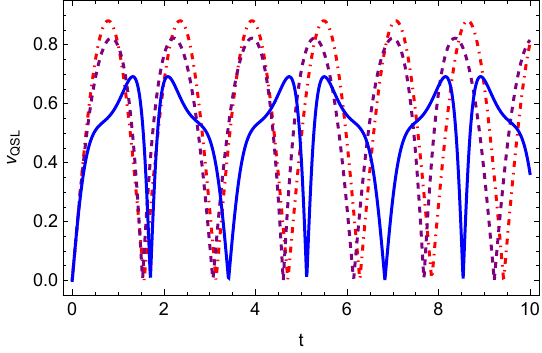}
\caption{\label{fig:qsl_thermo} Quantum speed limit \eqref{eq:qsl_qubit} for the nonlinear thermometer \eqref{eq:thermo} with a Gross-Pitaevskii nonlinearity, $\kappa(x)=g x^2$, (left panel) and a logarithmic nonlinearity, $\kappa(x)=g \ln(x^2)$, (right panel) with the initial state $\rho_0$ \eqref{eq:rho0}. Parameters are $\Delta=1$, $J=1$, $\beta=1$, for $g= 0$ (red, dotdashed line), $g=1$ (purple, dashed line), and $g= 2$ (blue, solid line).}
\end{figure*}

In Fig.~\ref{fig:qsl_thermo} we plot the resulting QSL \eqref{eq:vQSL} again for the Gross-Pitaevskii nonlinearity, $\kappa(x)=g x^2$, as well as the logarithmic nonlinearity, $\kappa(x)=g \ln(x^2)$. We observe that in contrast to single systems undergoing nonlinear dynamics the maximal QSL is actually reduced. This is an interesting consequence of the nonlinear evolution, since typically the QSL for open system dynamics is larger than for isolated dynamics \cite{Deffner2013PRL,Cimmarusti2015PRL}. However, as before in the isolated case we also observe that the effect is more pronounced for the logarithmic case.

\begin{figure*}
\includegraphics[width=.48\textwidth]{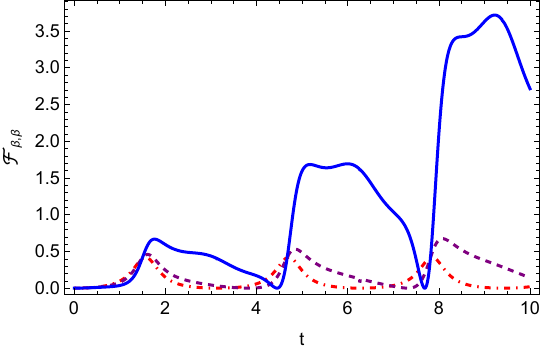}\hfill \includegraphics[width=.48\textwidth]{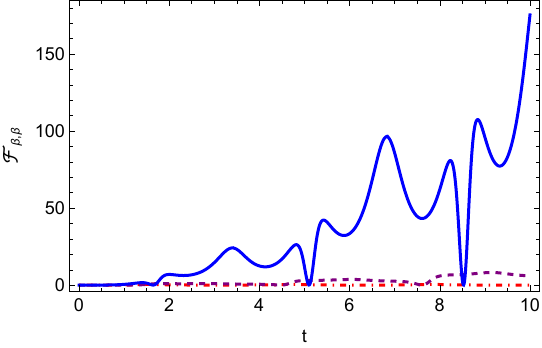}
\caption{\label{fig:fish_thermo} Quantum Fisher information for temperature, $\mc{F}_{\beta,\beta}$, \eqref{eq:fisher_qubit} for the nonlinear thermometer \eqref{eq:thermo} with a Gross-Pitaevskii nonlinearity, $\kappa(x)=g x^2$, (left panel) and a logarithmic nonlinearity, $\kappa(x)=g \ln(x^2)$, (right panel) with the initial state $\rho_0$ \eqref{eq:rho0}. Parameters are $\Delta=1$, $J=1$, $\beta=1$, for $g= 0$ (red, dotdashed line), $g=1$ (purple, dashed line), and $g= 2$ (blue, solid line).}
\end{figure*}

Despite the fact the QSL is reduced, the quantum Fisher information for temperature \eqref{eq:fisher_qubit} exhibits a similar enhancement for nonlinear thermometers, cf. Fig.~\ref{fig:fish_thermo}. In the two-qubit case this effect of the nonlinearity is significantly stronger than for the Landau-Zener model, cf. Fig.~\ref{fig:fish_LZ}.

\paragraph*{Discussion of the main results}

The time-dependent decrease of the QSL and the increase of the quantum Fisher information can be understood from the dynamics on the Bloch sphere of the thermometer qubit, $\mf{T}$. In the linear case, $\mf{T}$, simply oscillates along the $z$-axis. See Eq.~\eqref{eq:app_z} for an analytical expression. In the nonlinear case, also the $x$ and $y$ components of the Bloch vector are time-dependent. As has been analyzed in Ref.~\cite{Childs2016PRA}, the nonlinear term induces an additional flow around the Bloch sphere, and the magnitude of this flow only depends on the polar angle. Thus, the nonlinear term in the evolution equation can be understood as a ``drag force'' that hinders the free oscillation of the state along the $z$-axis. This leads to a decrease of the rate with which the time-dependent state becomes distinguishable from its history, and the QSL \eqref{eq:vQSL} is reduced. However, as a function of time the quantum states becomes increasingly more ``twisted'' around the Bloch sphere, which leads to a more ``structured'' state. Hence, the expected value of the extractable information, i.e., the quantum Fisher information, increases.

\section{Concluding remarks}

It had already been noted in the literature that nonlinear interactions in complex quantum many body systems can be leveraged to beat the standard Heisenberg limit  \cite{Boixo2007PRL,Roy2008PRL,Beau2017PRL}. In the present work, we have shown that nonlinear terms in the evolution equation can be exploited as a resource to improve the precision in quantum thermometry. To this end, we studied three example systems consisting of single and two qubits. Despite the simplicity of the models, we have been able to draw some general conclusions. Namely, the precision of thermometry, as quantified by the quantum Fisher information, can be enhanced if the thermal Gibbs state is not invariant under the dynamics. A good indicator for this possible enhancement is the quantum speed limit, which for isolated systems is boosted by the nonlinear evolution. More generally, the nonlinear term leads to additional flows around the Bloch sphere, which may decrease the quantum speed. However, in all considered cases we have found a stark enhancement of the quantum Fisher information, and the effect was more pronounced for logarithmic nonlinearities than for the Gross-Pitaevskii equation.

These results open the door to a whole host of questions. First and foremost, it would be interesting to study if and to what extent our present findings can be generalized to a proper description of Bose-Einstein condensates and Bose liquids. To this end, it would be important to study to what extent the nonlinearities can be designed to give optimal precision, and what can be experimentally realized. Such an avenue of research might eventually lead to the development of highly precise thermometers for ultra-cold temperatures.

\begin{acknowledgments}
It is a pleasure to thank the original quantum wizard, Steve Campbell, for many insightful discussions. S.D. acknowledges support from  the U.S. National Science Foundation under Grant No. DMR-2010127 and the John Templeton Foundation under Grant No. 62422. 
\end{acknowledgments}

\appendix

\section{\label{sec:appA} Solving the dynamics}

In this appendix, we collect mathematical details required to solve the dynamics of the nonlinear thermometer \eqref{eq:thermo}. The equations of motion for the two-qubit state \eqref{eq:2q} undergoing the nonlinear evolution described by Eq.~\eqref{eq:thermo} can be written as
\begin{equation}
\begin{aligned}[b]
\dot{\alpha}_{0,0}&= \quad 0 \\
\dot{\alpha}_{0,x}&=\quad 2 \Delta\,\alpha_{0,y}-J\,\alpha_{y,z} \\
\dot{\alpha}_{0,y}&=\, -\, 2 \Delta\,\alpha_{0,x}+ J\,\alpha_{x,z}\\ 
\dot{\alpha}_{0,z}&=\, -\, J \left(\alpha_{x,y}-\alpha_{y,x}\right)\,,
\end{aligned}
\end{equation}
where we suppressed the time-dependence of the coefficients to reduce clutter in the formulas. Note that this set is independent on the nonlinear term. However, the coefficients for $x$ and $y$ are governed by the nonlinear term,
\begin{equation}
\begin{aligned}[b]
\dot{\alpha}_{x,0}&=\quad\Delta\, \alpha_{y,0}\left[2+\tilde{\kappa}(\alpha_{z,0})\right]-J\, \alpha_{z,y} \\
\dot{\alpha}_{x,x}&= \quad \Delta\,\left\{2 \alpha_{x,y}+\alpha_{y,x}[2+\tilde{\kappa}(\alpha_{z,0})]\right\}\\
\dot{\alpha}_{x,y}&=\, - \, \Delta\,\left\{2 \alpha_{x,x}- \alpha_{y,y}[2+\tilde{\kappa}(\alpha_{z,0})]\right\}+J (\alpha_{0,z}-\alpha_{z,0}) \\
\dot{\alpha}_{x,z}&=\quad  \Delta \,\alpha_{y,z} [2+\tilde{\kappa}(\alpha_{z,0})]-J \alpha_{0,y}
\end{aligned}
\end{equation}
and
\begin{equation}
\begin{aligned}[b]
\dot{\alpha}_{y,0}&=\, -\,\Delta \alpha_{x,0} [2+\tilde{\kappa}(\alpha_{z,0})]+J \,\alpha_{z,x}\\
\dot{\alpha}_{y,x}&=\, -\,\Delta\,\left\{2 \alpha_{x,x} [2+\tilde{\kappa}(\alpha_{z,0})]- \alpha_{y,y} \right\}-J (\alpha_{0,z}-\alpha_{z,0})\\
\dot{\alpha}_{y,y}&=\, -\,\Delta \left\{2 \alpha_{y,x}+2\alpha_{x,y}[2+\tilde{\kappa}(\alpha_{z,0})]\right\}\\
\dot{\alpha}_{y,z}&=\, -\,\Delta \,\alpha_{x,z} [2+\tilde{\kappa}(\alpha_{z,0})]+J \alpha_{0,x}\,.
\end{aligned}
\end{equation}
Finally, the coefficients for the $z$-component of the thermometer are again independent of the nonlinear term,
\begin{equation}
\begin{aligned}[b]
\dot{\alpha}_{z,0}&= \quad J \left(\alpha_{x,y}-\alpha_{y,x}\right)\\
\dot{\alpha}_{z,x}&= \quad 2\Delta\,\alpha_{z,y}-J \alpha_{y,0}\\
\dot{\alpha}_{z,y}&=\, -\, 2\Delta\,\alpha_{z,x}+J\,\alpha_{x,0}\\
\dot{\alpha}_{z,z}&=\quad 0\,.
\end{aligned}
\end{equation}

\paragraph*{Linear dynamics}

Interestingly, the dynamics can be solved analytically for $\tilde{\kappa}(\alpha_{z,0})=0$. We have,
\begin{equation}
\begin{aligned}[b]
\alpha_{0,0}&= 1 \\
\alpha_{0,x}&= \alpha_{0,y}= 0 \\
\alpha_{0,z}&=\cos^2(J t)- \sin^2(J t)\,\tah{\beta\Delta}\,,
\end{aligned}
\end{equation}
and
\begin{equation}
\begin{aligned}[b]
\alpha_{x,0}&=\alpha_{x,x}=\alpha_{x,z}=0 \\
\alpha_{x,y}&=\frac{1}{2}\si{2 J t}\left[1+\tah{\beta \Delta}\right]\,.
\end{aligned}
\end{equation}
Similarly, we find
\begin{equation}
\begin{aligned}[b]
\alpha_{y,0}&=\alpha_{y,y}=\alpha_{y,z}=0\\
\alpha_{y,x}&=-\frac{1}{2}\si{2 J t}\left[1+\tah{\beta \Delta}\right]\,,
\end{aligned}
\end{equation}
and also
\begin{equation}
\begin{aligned}[b]
\alpha_{z,0}&= \sin^2(J t)-\cos^2(J t) \tah{\beta \Delta}\\
\alpha_{z,x}&= \alpha_{z,y}=0\\
\alpha_{z,z}&=-\tah{\beta \Delta}\,.
\end{aligned}
\end{equation}
Note that this means that the time-dependent state of the thermometer qubit \eqref{eq:T} simply reads,
\begin{equation}
\label{eq:app_z}
\rho_t^\mf{T}=\frac{1}{2}\left[\id_2 +\left(\sin^2(J t)-\cos^2(J t) \tah{\beta \Delta}\right)\,\sigma_z\right]\,.
\end{equation}
Hence, the quantum state oscillates along the $z$-axis inside the Bloch sphere. Moreover, we also obtain a closed form expression for the QSL \eqref{eq:qsl_qubit}
\begin{equation}
v_\mrm{QSL}=\left|J \co{Jt}\si{Jt}\left[1+\tah{\beta \Delta}\right]\right|
\end{equation}
as well as temperature component of the for quantum Fisher information
\begin{widetext}
\begin{equation}
\mc{F}_{\beta,\beta}=\frac{\Delta ^2 \text{sech}^4(\beta  \Delta ) \cos ^4(J t)}{1-\tanh (\beta  \Delta ) \cos ^2(J t)
   \left[\tanh (\beta  \Delta ) \cos ^2(J t)+\cos (2 J t)-1\right]-\sin ^4(J t)}\,.
\end{equation}
\end{widetext}

\bibliography{nonlinear_thermo}

\begin{thebibliography}{72}%
\makeatletter
\providecommand \@ifxundefined [1]{%
 \@ifx{#1\undefined}
}%
\providecommand \@ifnum [1]{%
 \ifnum #1\expandafter \@firstoftwo
 \else \expandafter \@secondoftwo
 \fi
}%
\providecommand \@ifx [1]{%
 \ifx #1\expandafter \@firstoftwo
 \else \expandafter \@secondoftwo
 \fi
}%
\providecommand \natexlab [1]{#1}%
\providecommand \enquote  [1]{``#1''}%
\providecommand \bibnamefont  [1]{#1}%
\providecommand \bibfnamefont [1]{#1}%
\providecommand \citenamefont [1]{#1}%
\providecommand \href@noop [0]{\@secondoftwo}%
\providecommand \href [0]{\begingroup \@sanitize@url \@href}%
\providecommand \@href[1]{\@@startlink{#1}\@@href}%
\providecommand \@@href[1]{\endgroup#1\@@endlink}%
\providecommand \@sanitize@url [0]{\catcode `\\12\catcode `\$12\catcode
  `\&12\catcode `\#12\catcode `\^12\catcode `\_12\catcode `\%12\relax}%
\providecommand \@@startlink[1]{}%
\providecommand \@@endlink[0]{}%
\providecommand \url  [0]{\begingroup\@sanitize@url \@url }%
\providecommand \@url [1]{\endgroup\@href {#1}{\urlprefix }}%
\providecommand \urlprefix  [0]{URL }%
\providecommand \Eprint [0]{\href }%
\providecommand \doibase [0]{https://doi.org/}%
\providecommand \selectlanguage [0]{\@gobble}%
\providecommand \bibinfo  [0]{\@secondoftwo}%
\providecommand \bibfield  [0]{\@secondoftwo}%
\providecommand \translation [1]{[#1]}%
\providecommand \BibitemOpen [0]{}%
\providecommand \bibitemStop [0]{}%
\providecommand \bibitemNoStop [0]{.\EOS\space}%
\providecommand \EOS [0]{\spacefactor3000\relax}%
\providecommand \BibitemShut  [1]{\csname bibitem#1\endcsname}%
\let\auto@bib@innerbib\@empty
\bibitem [{\citenamefont {Carnot}(1824)}]{Carnot1824}%
  \BibitemOpen
  \bibfield  {author} {\bibinfo {author} {\bibfnamefont {S.}~\bibnamefont
  {Carnot}},\ }\href@noop {} {\emph {\bibinfo {title} {Reflections on the
  motive power of fire, and on machines fitted to develop that power}}}\
  (\bibinfo  {publisher} {Paris: Bachelier},\ \bibinfo {year}
  {1824})\BibitemShut {NoStop}%
\bibitem [{\citenamefont {Callen}(1985)}]{Callen1985}%
  \BibitemOpen
  \bibfield  {author} {\bibinfo {author} {\bibfnamefont {H.~B.}\ \bibnamefont
  {Callen}},\ }\href@noop {} {\emph {\bibinfo {title} {Thermodynamics and an
  introduction to thermostatistics}}}\ (\bibinfo  {publisher} {Wiley},\
  \bibinfo {address} {New York, USA},\ \bibinfo {year} {1985})\BibitemShut
  {NoStop}%
\bibitem [{\citenamefont {Fretwell}(1937)}]{fretwell1937development}%
  \BibitemOpen
  \bibfield  {author} {\bibinfo {author} {\bibfnamefont {M.~B.}\ \bibnamefont
  {Fretwell}},\ }\bibfield  {title} {\bibinfo {title} {The development of the
  thermometer},\ }\href {https://doi.org/10.5951/MT.30.2.0080} {\bibfield
  {journal} {\bibinfo  {journal} {The Mathematics Teacher}\ }\textbf {\bibinfo
  {volume} {30}},\ \bibinfo {pages} {80} (\bibinfo {year} {1937})}\BibitemShut
  {NoStop}%
\bibitem [{\citenamefont {Tresca}\ \emph {et~al.}(2022)\citenamefont {Tresca},
  \citenamefont {Profeta}, \citenamefont {Marini}, \citenamefont {Bachelet},
  \citenamefont {Sanna}, \citenamefont {Calandra},\ and\ \citenamefont
  {Boeri}}]{Tresca2022PRB}%
  \BibitemOpen
  \bibfield  {author} {\bibinfo {author} {\bibfnamefont {C.}~\bibnamefont
  {Tresca}}, \bibinfo {author} {\bibfnamefont {G.}~\bibnamefont {Profeta}},
  \bibinfo {author} {\bibfnamefont {G.}~\bibnamefont {Marini}}, \bibinfo
  {author} {\bibfnamefont {G.~B.}\ \bibnamefont {Bachelet}}, \bibinfo {author}
  {\bibfnamefont {A.}~\bibnamefont {Sanna}}, \bibinfo {author} {\bibfnamefont
  {M.}~\bibnamefont {Calandra}},\ and\ \bibinfo {author} {\bibfnamefont
  {L.}~\bibnamefont {Boeri}},\ }\bibfield  {title} {\bibinfo {title} {Why
  mercury is a superconductor},\ }\href
  {https://doi.org/10.1103/PhysRevB.106.L180501} {\bibfield  {journal}
  {\bibinfo  {journal} {Phys. Rev. B}\ }\textbf {\bibinfo {volume} {106}},\
  \bibinfo {pages} {L180501} (\bibinfo {year} {2022})}\BibitemShut {NoStop}%
\bibitem [{\citenamefont {Ashcroft}\ and\ \citenamefont
  {Mermin}(1976)}]{Ashcroft1976}%
  \BibitemOpen
  \bibfield  {author} {\bibinfo {author} {\bibfnamefont {N.~W.}\ \bibnamefont
  {Ashcroft}}\ and\ \bibinfo {author} {\bibfnamefont {N.~D.}\ \bibnamefont
  {Mermin}},\ }\href@noop {} {\emph {\bibinfo {title} {Solid State Physics}}}\
  (\bibinfo  {publisher} {Thomson Learning, Inc.},\ \bibinfo {year}
  {1976})\BibitemShut {NoStop}%
\bibitem [{\citenamefont {Huang}(2009)}]{huang2009introduction}%
  \BibitemOpen
  \bibfield  {author} {\bibinfo {author} {\bibfnamefont {K.}~\bibnamefont
  {Huang}},\ }\href@noop {} {\emph {\bibinfo {title} {Introduction to
  statistical physics}}}\ (\bibinfo  {publisher} {Chapman and Hall/CRC},\
  \bibinfo {year} {2009})\BibitemShut {NoStop}%
\bibitem [{\citenamefont {Deffner}(2022)}]{Deffner2022EPL}%
  \BibitemOpen
  \bibfield  {author} {\bibinfo {author} {\bibfnamefont {S.}~\bibnamefont
  {Deffner}},\ }\bibfield  {title} {\bibinfo {title} {Nonlinear speed-ups in
  ultracold quantum gases},\ }\href {https://doi.org/10.1209/0295-5075/ac9fed}
  {\bibfield  {journal} {\bibinfo  {journal} {EPL (Europhys. Lett.)}\ }\textbf
  {\bibinfo {volume} {140}},\ \bibinfo {pages} {48001} (\bibinfo {year}
  {2022})}\BibitemShut {NoStop}%
\bibitem [{\citenamefont {Poggi}\ \emph {et~al.}(2013)\citenamefont {Poggi},
  \citenamefont {Lombardo},\ and\ \citenamefont {Wisniacki}}]{Poggi2013EPL}%
  \BibitemOpen
  \bibfield  {author} {\bibinfo {author} {\bibfnamefont {P.~M.}\ \bibnamefont
  {Poggi}}, \bibinfo {author} {\bibfnamefont {F.~C.}\ \bibnamefont
  {Lombardo}},\ and\ \bibinfo {author} {\bibfnamefont {D.~A.}\ \bibnamefont
  {Wisniacki}},\ }\bibfield  {title} {\bibinfo {title} {Quantum speed limit and
  optimal evolution time in a two-level system},\ }\href
  {https://doi.org/10.1209/0295-5075/104/40005} {\bibfield  {journal} {\bibinfo
   {journal} {{EPL} (Europhys. Lett.)}\ }\textbf {\bibinfo {volume} {104}},\
  \bibinfo {pages} {40005} (\bibinfo {year} {2013})}\BibitemShut {NoStop}%
\bibitem [{\citenamefont {Poggi}(2019)}]{Poggi2016PRA}%
  \BibitemOpen
  \bibfield  {author} {\bibinfo {author} {\bibfnamefont {P.~M.}\ \bibnamefont
  {Poggi}},\ }\bibfield  {title} {\bibinfo {title} {Geometric quantum speed
  limits and short-time accessibility to unitary operations},\ }\href
  {https://doi.org/doi.org/10.1103/PhysRevA.99.042116} {\bibfield  {journal}
  {\bibinfo  {journal} {Phys. Rev. A}\ }\textbf {\bibinfo {volume} {99}},\
  \bibinfo {pages} {042116} (\bibinfo {year} {2019})}\BibitemShut {NoStop}%
\bibitem [{\citenamefont {Deffner}\ and\ \citenamefont
  {Campbell}(2017)}]{Deffner2017JPA}%
  \BibitemOpen
  \bibfield  {author} {\bibinfo {author} {\bibfnamefont {S.}~\bibnamefont
  {Deffner}}\ and\ \bibinfo {author} {\bibfnamefont {S.}~\bibnamefont
  {Campbell}},\ }\bibfield  {title} {\bibinfo {title} {{Quantum speed limits:
  from {Heisenberg}'s uncertainty principle to optimal quantum control}},\
  }\href {https://doi.org/10.1088/1751-8121/aa86c6} {\bibfield  {journal}
  {\bibinfo  {journal} {J. Phys. A: Math. Theor.}\ }\textbf {\bibinfo {volume}
  {50}},\ \bibinfo {pages} {453001} (\bibinfo {year} {2017})}\BibitemShut
  {NoStop}%
\bibitem [{\citenamefont {Giovannetti}\ \emph {et~al.}(2011)\citenamefont
  {Giovannetti}, \citenamefont {Lloyd},\ and\ \citenamefont
  {Maccone}}]{Giovannetti2011NPho}%
  \BibitemOpen
  \bibfield  {author} {\bibinfo {author} {\bibfnamefont {V.}~\bibnamefont
  {Giovannetti}}, \bibinfo {author} {\bibfnamefont {S.}~\bibnamefont {Lloyd}},\
  and\ \bibinfo {author} {\bibfnamefont {L.}~\bibnamefont {Maccone}},\
  }\bibfield  {title} {\bibinfo {title} {Advances in quantum metrology},\
  }\href {https://doi.org/10.1038/nphoton.2011.35} {\bibfield  {journal}
  {\bibinfo  {journal} {Nature Photon.}\ }\textbf {\bibinfo {volume} {5}},\
  \bibinfo {pages} {222} (\bibinfo {year} {2011})}\BibitemShut {NoStop}%
\bibitem [{\citenamefont {Boixo}\ \emph {et~al.}(2007)\citenamefont {Boixo},
  \citenamefont {Flammia}, \citenamefont {Caves},\ and\ \citenamefont
  {Geremia}}]{Boixo2007PRL}%
  \BibitemOpen
  \bibfield  {author} {\bibinfo {author} {\bibfnamefont {S.}~\bibnamefont
  {Boixo}}, \bibinfo {author} {\bibfnamefont {S.~T.}\ \bibnamefont {Flammia}},
  \bibinfo {author} {\bibfnamefont {C.~M.}\ \bibnamefont {Caves}},\ and\
  \bibinfo {author} {\bibfnamefont {J.}~\bibnamefont {Geremia}},\ }\bibfield
  {title} {\bibinfo {title} {Generalized limits for single-parameter quantum
  estimation},\ }\href {https://doi.org/10.1103/PhysRevLett.98.090401}
  {\bibfield  {journal} {\bibinfo  {journal} {Phys. Rev. Lett.}\ }\textbf
  {\bibinfo {volume} {98}},\ \bibinfo {pages} {090401} (\bibinfo {year}
  {2007})}\BibitemShut {NoStop}%
\bibitem [{\citenamefont {Roy}\ and\ \citenamefont
  {Braunstein}(2008)}]{Roy2008PRL}%
  \BibitemOpen
  \bibfield  {author} {\bibinfo {author} {\bibfnamefont {S.~M.}\ \bibnamefont
  {Roy}}\ and\ \bibinfo {author} {\bibfnamefont {S.~L.}\ \bibnamefont
  {Braunstein}},\ }\bibfield  {title} {\bibinfo {title} {Exponentially enhanced
  quantum metrology},\ }\href {https://doi.org/10.1103/PhysRevLett.100.220501}
  {\bibfield  {journal} {\bibinfo  {journal} {Phys. Rev. Lett.}\ }\textbf
  {\bibinfo {volume} {100}},\ \bibinfo {pages} {220501} (\bibinfo {year}
  {2008})}\BibitemShut {NoStop}%
\bibitem [{\citenamefont {Beau}\ and\ \citenamefont {del
  Campo}(2017)}]{Beau2017PRL}%
  \BibitemOpen
  \bibfield  {author} {\bibinfo {author} {\bibfnamefont {M.}~\bibnamefont
  {Beau}}\ and\ \bibinfo {author} {\bibfnamefont {A.}~\bibnamefont {del
  Campo}},\ }\bibfield  {title} {\bibinfo {title} {Nonlinear quantum metrology
  of many-body open systems},\ }\href
  {https://doi.org/10.1103/PhysRevLett.119.010403} {\bibfield  {journal}
  {\bibinfo  {journal} {Phys. Rev. Lett.}\ }\textbf {\bibinfo {volume} {119}},\
  \bibinfo {pages} {010403} (\bibinfo {year} {2017})}\BibitemShut {NoStop}%
\bibitem [{\citenamefont {Yang}\ \emph {et~al.}(2022)\citenamefont {Yang},
  \citenamefont {Pang}, \citenamefont {del Campo},\ and\ \citenamefont
  {Jordan}}]{Yang2022PRR}%
  \BibitemOpen
  \bibfield  {author} {\bibinfo {author} {\bibfnamefont {J.}~\bibnamefont
  {Yang}}, \bibinfo {author} {\bibfnamefont {S.}~\bibnamefont {Pang}}, \bibinfo
  {author} {\bibfnamefont {A.}~\bibnamefont {del Campo}},\ and\ \bibinfo
  {author} {\bibfnamefont {A.~N.}\ \bibnamefont {Jordan}},\ }\bibfield  {title}
  {\bibinfo {title} {Super-heisenberg scaling in hamiltonian parameter
  estimation in the long-range kitaev chain},\ }\href
  {https://doi.org/10.1103/PhysRevResearch.4.013133} {\bibfield  {journal}
  {\bibinfo  {journal} {Phys. Rev. Res.}\ }\textbf {\bibinfo {volume} {4}},\
  \bibinfo {pages} {013133} (\bibinfo {year} {2022})}\BibitemShut {NoStop}%
\bibitem [{\citenamefont {Campbell}\ \emph {et~al.}(2018)\citenamefont
  {Campbell}, \citenamefont {Genoni},\ and\ \citenamefont
  {Deffner}}]{Campbell2018QST}%
  \BibitemOpen
  \bibfield  {author} {\bibinfo {author} {\bibfnamefont {S.}~\bibnamefont
  {Campbell}}, \bibinfo {author} {\bibfnamefont {M.~G.}\ \bibnamefont
  {Genoni}},\ and\ \bibinfo {author} {\bibfnamefont {S.}~\bibnamefont
  {Deffner}},\ }\bibfield  {title} {\bibinfo {title} {Precision thermometry and
  the quantum speed limit},\ }\href {https://doi.org/10.1088/2058-9565/aaa641}
  {\bibfield  {journal} {\bibinfo  {journal} {Quantum Sci. Technol.}\ }\textbf
  {\bibinfo {volume} {3}},\ \bibinfo {pages} {025002} (\bibinfo {year}
  {2018})}\BibitemShut {NoStop}%
\bibitem [{\citenamefont {Mehboudi}\ \emph {et~al.}(2019)\citenamefont
  {Mehboudi}, \citenamefont {Sanpera},\ and\ \citenamefont
  {Correa}}]{Mehboudi2019JPA}%
  \BibitemOpen
  \bibfield  {author} {\bibinfo {author} {\bibfnamefont {M.}~\bibnamefont
  {Mehboudi}}, \bibinfo {author} {\bibfnamefont {A.}~\bibnamefont {Sanpera}},\
  and\ \bibinfo {author} {\bibfnamefont {L.~A.}\ \bibnamefont {Correa}},\
  }\bibfield  {title} {\bibinfo {title} {Thermometry in the quantum regime:
  recent theoretical progress},\ }\href
  {https://doi.org/10.1088/1751-8121/ab2828} {\bibfield  {journal} {\bibinfo
  {journal} {J. Phys. A: Math. Theor.}\ }\textbf {\bibinfo {volume} {52}},\
  \bibinfo {pages} {303001} (\bibinfo {year} {2019})}\BibitemShut {NoStop}%
\bibitem [{\citenamefont {Campbell}\ \emph {et~al.}(2017)\citenamefont
  {Campbell}, \citenamefont {Mehboudi}, \citenamefont {Chiara},\ and\
  \citenamefont {Paternostro}}]{Campbell2017NJP}%
  \BibitemOpen
  \bibfield  {author} {\bibinfo {author} {\bibfnamefont {S.}~\bibnamefont
  {Campbell}}, \bibinfo {author} {\bibfnamefont {M.}~\bibnamefont {Mehboudi}},
  \bibinfo {author} {\bibfnamefont {G.~D.}\ \bibnamefont {Chiara}},\ and\
  \bibinfo {author} {\bibfnamefont {M.}~\bibnamefont {Paternostro}},\
  }\bibfield  {title} {\bibinfo {title} {Global and local thermometry schemes
  in coupled quantum systems},\ }\href
  {https://doi.org/10.1088/1367-2630/aa7fac} {\bibfield  {journal} {\bibinfo
  {journal} {New J. Phys.}\ }\textbf {\bibinfo {volume} {19}},\ \bibinfo
  {pages} {103003} (\bibinfo {year} {2017})}\BibitemShut {NoStop}%
\bibitem [{\citenamefont {Kiilerich}\ \emph {et~al.}(2018)\citenamefont
  {Kiilerich}, \citenamefont {De~Pasquale},\ and\ \citenamefont
  {Giovannetti}}]{Kiilerich2018PRA}%
  \BibitemOpen
  \bibfield  {author} {\bibinfo {author} {\bibfnamefont {A.~H.}\ \bibnamefont
  {Kiilerich}}, \bibinfo {author} {\bibfnamefont {A.}~\bibnamefont
  {De~Pasquale}},\ and\ \bibinfo {author} {\bibfnamefont {V.}~\bibnamefont
  {Giovannetti}},\ }\bibfield  {title} {\bibinfo {title} {Dynamical approach to
  ancilla-assisted quantum thermometry},\ }\href
  {https://doi.org/10.1103/PhysRevA.98.042124} {\bibfield  {journal} {\bibinfo
  {journal} {Phys. Rev. A}\ }\textbf {\bibinfo {volume} {98}},\ \bibinfo
  {pages} {042124} (\bibinfo {year} {2018})}\BibitemShut {NoStop}%
\bibitem [{\citenamefont {Seveso}\ and\ \citenamefont
  {Paris}(2018)}]{Seveso2018PRA}%
  \BibitemOpen
  \bibfield  {author} {\bibinfo {author} {\bibfnamefont {L.}~\bibnamefont
  {Seveso}}\ and\ \bibinfo {author} {\bibfnamefont {M.~G.~A.}\ \bibnamefont
  {Paris}},\ }\bibfield  {title} {\bibinfo {title} {Trade-off between
  information and disturbance in qubit thermometry},\ }\href
  {https://doi.org/10.1103/PhysRevA.97.032129} {\bibfield  {journal} {\bibinfo
  {journal} {Phys. Rev. A}\ }\textbf {\bibinfo {volume} {97}},\ \bibinfo
  {pages} {032129} (\bibinfo {year} {2018})}\BibitemShut {NoStop}%
\bibitem [{\citenamefont {{Razavian, Sholeh}}\ \emph
  {et~al.}(2019)\citenamefont {{Razavian, Sholeh}}, \citenamefont {{Benedetti,
  Claudia}}, \citenamefont {{Bina, Matteo}}, \citenamefont {{Akbari-Kourbolagh,
  Yahya}},\ and\ \citenamefont {{Paris, Matteo G. A.}}}]{Razavian2019}%
  \BibitemOpen
  \bibfield  {author} {\bibinfo {author} {\bibnamefont {{Razavian, Sholeh}}},
  \bibinfo {author} {\bibnamefont {{Benedetti, Claudia}}}, \bibinfo {author}
  {\bibnamefont {{Bina, Matteo}}}, \bibinfo {author} {\bibnamefont
  {{Akbari-Kourbolagh, Yahya}}},\ and\ \bibinfo {author} {\bibnamefont {{Paris,
  Matteo G. A.}}},\ }\bibfield  {title} {\bibinfo {title} {Quantum thermometry
  by single-qubit dephasing},\ }\href
  {https://doi.org/10.1140/epjp/i2019-12708-9} {\bibfield  {journal} {\bibinfo
  {journal} {Eur. Phys. J. Plus}\ }\textbf {\bibinfo {volume} {134}},\ \bibinfo
  {pages} {284} (\bibinfo {year} {2019})}\BibitemShut {NoStop}%
\bibitem [{\citenamefont {Mukherjee}\ \emph {et~al.}(2019)\citenamefont
  {Mukherjee}, \citenamefont {Zwick}, \citenamefont {Ghosh}, \citenamefont
  {Chen},\ and\ \citenamefont {Kurizki}}]{Mukherjee2019CP}%
  \BibitemOpen
  \bibfield  {author} {\bibinfo {author} {\bibfnamefont {V.}~\bibnamefont
  {Mukherjee}}, \bibinfo {author} {\bibfnamefont {A.}~\bibnamefont {Zwick}},
  \bibinfo {author} {\bibfnamefont {A.}~\bibnamefont {Ghosh}}, \bibinfo
  {author} {\bibfnamefont {X.}~\bibnamefont {Chen}},\ and\ \bibinfo {author}
  {\bibfnamefont {G.}~\bibnamefont {Kurizki}},\ }\bibfield  {title} {\bibinfo
  {title} {Enhanced precision bound of low-temperature quantum thermometry via
  dynamical control},\ }\href {https://doi.org/10.1038/s42005-019-0265-y}
  {\bibfield  {journal} {\bibinfo  {journal} {Commun. Phys.}\ }\textbf
  {\bibinfo {volume} {2}},\ \bibinfo {pages} {162} (\bibinfo {year}
  {2019})}\BibitemShut {NoStop}%
\bibitem [{\citenamefont {Feyles}\ \emph {et~al.}(2019)\citenamefont {Feyles},
  \citenamefont {Mancino}, \citenamefont {Sbroscia}, \citenamefont {Gianani},\
  and\ \citenamefont {Barbieri}}]{Feyles2019PRA}%
  \BibitemOpen
  \bibfield  {author} {\bibinfo {author} {\bibfnamefont {M.~M.}\ \bibnamefont
  {Feyles}}, \bibinfo {author} {\bibfnamefont {L.}~\bibnamefont {Mancino}},
  \bibinfo {author} {\bibfnamefont {M.}~\bibnamefont {Sbroscia}}, \bibinfo
  {author} {\bibfnamefont {I.}~\bibnamefont {Gianani}},\ and\ \bibinfo {author}
  {\bibfnamefont {M.}~\bibnamefont {Barbieri}},\ }\bibfield  {title} {\bibinfo
  {title} {Dynamical role of quantum signatures in quantum thermometry},\
  }\href {https://doi.org/10.1103/PhysRevA.99.062114} {\bibfield  {journal}
  {\bibinfo  {journal} {Phys. Rev. A}\ }\textbf {\bibinfo {volume} {99}},\
  \bibinfo {pages} {062114} (\bibinfo {year} {2019})}\BibitemShut {NoStop}%
\bibitem [{\citenamefont {Mitchison}\ \emph {et~al.}(2020)\citenamefont
  {Mitchison}, \citenamefont {Fogarty}, \citenamefont {Guarnieri},
  \citenamefont {Campbell}, \citenamefont {Busch},\ and\ \citenamefont
  {Goold}}]{Mitchison2020PRL}%
  \BibitemOpen
  \bibfield  {author} {\bibinfo {author} {\bibfnamefont {M.~T.}\ \bibnamefont
  {Mitchison}}, \bibinfo {author} {\bibfnamefont {T.}~\bibnamefont {Fogarty}},
  \bibinfo {author} {\bibfnamefont {G.}~\bibnamefont {Guarnieri}}, \bibinfo
  {author} {\bibfnamefont {S.}~\bibnamefont {Campbell}}, \bibinfo {author}
  {\bibfnamefont {T.}~\bibnamefont {Busch}},\ and\ \bibinfo {author}
  {\bibfnamefont {J.}~\bibnamefont {Goold}},\ }\bibfield  {title} {\bibinfo
  {title} {In situ thermometry of a cold fermi gas via dephasing impurities},\
  }\href {https://doi.org/10.1103/PhysRevLett.125.080402} {\bibfield  {journal}
  {\bibinfo  {journal} {Phys. Rev. Lett.}\ }\textbf {\bibinfo {volume} {125}},\
  \bibinfo {pages} {080402} (\bibinfo {year} {2020})}\BibitemShut {NoStop}%
\bibitem [{\citenamefont {O’Connor}\ \emph {et~al.}(2021)\citenamefont
  {O’Connor}, \citenamefont {Vacchini},\ and\ \citenamefont
  {Campbell}}]{Connor2012Entropy}%
  \BibitemOpen
  \bibfield  {author} {\bibinfo {author} {\bibfnamefont {E.}~\bibnamefont
  {O’Connor}}, \bibinfo {author} {\bibfnamefont {B.}~\bibnamefont
  {Vacchini}},\ and\ \bibinfo {author} {\bibfnamefont {S.}~\bibnamefont
  {Campbell}},\ }\bibfield  {title} {\bibinfo {title} {Stochastic collisional
  quantum thermometry},\ }\href {https://doi.org/10.3390/e23121634} {\bibfield
  {journal} {\bibinfo  {journal} {Entropy}\ }\textbf {\bibinfo {volume} {23}},\
  \bibinfo {pages} {1634} (\bibinfo {year} {2021})}\BibitemShut {NoStop}%
\bibitem [{\citenamefont {Hovhannisyan}\ \emph {et~al.}(2021)\citenamefont
  {Hovhannisyan}, \citenamefont {J\o{}rgensen}, \citenamefont {Landi},
  \citenamefont {Alhambra}, \citenamefont {Brask},\ and\ \citenamefont
  {Perarnau-Llobet}}]{HovhannisyanPRXQ2021}%
  \BibitemOpen
  \bibfield  {author} {\bibinfo {author} {\bibfnamefont {K.~V.}\ \bibnamefont
  {Hovhannisyan}}, \bibinfo {author} {\bibfnamefont {M.~R.}\ \bibnamefont
  {J\o{}rgensen}}, \bibinfo {author} {\bibfnamefont {G.~T.}\ \bibnamefont
  {Landi}}, \bibinfo {author} {\bibfnamefont {A.~M.}\ \bibnamefont {Alhambra}},
  \bibinfo {author} {\bibfnamefont {J.~B.}\ \bibnamefont {Brask}},\ and\
  \bibinfo {author} {\bibfnamefont {M.}~\bibnamefont {Perarnau-Llobet}},\
  }\bibfield  {title} {\bibinfo {title} {Optimal quantum thermometry with
  coarse-grained measurements},\ }\href
  {https://doi.org/10.1103/PRXQuantum.2.020322} {\bibfield  {journal} {\bibinfo
   {journal} {PRX Quantum}\ }\textbf {\bibinfo {volume} {2}},\ \bibinfo {pages}
  {020322} (\bibinfo {year} {2021})}\BibitemShut {NoStop}%
\bibitem [{\citenamefont {Mok}\ \emph {et~al.}(2021)\citenamefont {Mok},
  \citenamefont {Bharti}, \citenamefont {Kwek},\ and\ \citenamefont
  {Bayat}}]{Mok2021CP}%
  \BibitemOpen
  \bibfield  {author} {\bibinfo {author} {\bibfnamefont {W.-K.}\ \bibnamefont
  {Mok}}, \bibinfo {author} {\bibfnamefont {K.}~\bibnamefont {Bharti}},
  \bibinfo {author} {\bibfnamefont {L.-C.}\ \bibnamefont {Kwek}},\ and\
  \bibinfo {author} {\bibfnamefont {A.}~\bibnamefont {Bayat}},\ }\bibfield
  {title} {\bibinfo {title} {Optimal probes for global quantum thermometry},\
  }\href {https://doi.org/10.1038/s42005-021-00572-w} {\bibfield  {journal}
  {\bibinfo  {journal} {Commun. Phys.}\ }\textbf {\bibinfo {volume} {4}},\
  \bibinfo {pages} {62} (\bibinfo {year} {2021})}\BibitemShut {NoStop}%
\bibitem [{\citenamefont {Sekatski}\ and\ \citenamefont
  {Perarnau-Llobet}(2022)}]{Sekatski2022optimal}%
  \BibitemOpen
  \bibfield  {author} {\bibinfo {author} {\bibfnamefont {P.}~\bibnamefont
  {Sekatski}}\ and\ \bibinfo {author} {\bibfnamefont {M.}~\bibnamefont
  {Perarnau-Llobet}},\ }\bibfield  {title} {\bibinfo {title} {Optimal
  nonequilibrium thermometry in {M}arkovian environments},\ }\href
  {https://doi.org/10.22331/q-2022-12-07-869} {\bibfield  {journal} {\bibinfo
  {journal} {{Quantum}}\ }\textbf {\bibinfo {volume} {6}},\ \bibinfo {pages}
  {869} (\bibinfo {year} {2022})}\BibitemShut {NoStop}%
\bibitem [{\citenamefont {Albarelli}\ \emph {et~al.}(2023)\citenamefont
  {Albarelli}, \citenamefont {Paris}, \citenamefont {Vacchini},\ and\
  \citenamefont {Smirne}}]{Albarelli2023PRA}%
  \BibitemOpen
  \bibfield  {author} {\bibinfo {author} {\bibfnamefont {F.}~\bibnamefont
  {Albarelli}}, \bibinfo {author} {\bibfnamefont {M.~G.~A.}\ \bibnamefont
  {Paris}}, \bibinfo {author} {\bibfnamefont {B.}~\bibnamefont {Vacchini}},\
  and\ \bibinfo {author} {\bibfnamefont {A.}~\bibnamefont {Smirne}},\
  }\bibfield  {title} {\bibinfo {title} {Invasiveness of nonequilibrium
  pure-dephasing quantum thermometry},\ }\href
  {https://doi.org/10.1103/PhysRevA.108.062421} {\bibfield  {journal} {\bibinfo
   {journal} {Phys. Rev. A}\ }\textbf {\bibinfo {volume} {108}},\ \bibinfo
  {pages} {062421} (\bibinfo {year} {2023})}\BibitemShut {NoStop}%
\bibitem [{\citenamefont {Mihailescu}\ \emph {et~al.}(2023)\citenamefont
  {Mihailescu}, \citenamefont {Campbell},\ and\ \citenamefont
  {Mitchell}}]{Mihailescu2023PRA}%
  \BibitemOpen
  \bibfield  {author} {\bibinfo {author} {\bibfnamefont {G.}~\bibnamefont
  {Mihailescu}}, \bibinfo {author} {\bibfnamefont {S.}~\bibnamefont
  {Campbell}},\ and\ \bibinfo {author} {\bibfnamefont {A.~K.}\ \bibnamefont
  {Mitchell}},\ }\bibfield  {title} {\bibinfo {title} {Thermometry of strongly
  correlated fermionic quantum systems using impurity probes},\ }\href
  {https://doi.org/10.1103/PhysRevA.107.042614} {\bibfield  {journal} {\bibinfo
   {journal} {Phys. Rev. A}\ }\textbf {\bibinfo {volume} {107}},\ \bibinfo
  {pages} {042614} (\bibinfo {year} {2023})}\BibitemShut {NoStop}%
\bibitem [{\citenamefont {Mihailescu}\ \emph
  {et~al.}(2024{\natexlab{a}})\citenamefont {Mihailescu}, \citenamefont
  {Bayat}, \citenamefont {Campbell},\ and\ \citenamefont
  {Mitchell}}]{Mihailescu2024QST}%
  \BibitemOpen
  \bibfield  {author} {\bibinfo {author} {\bibfnamefont {G.}~\bibnamefont
  {Mihailescu}}, \bibinfo {author} {\bibfnamefont {A.}~\bibnamefont {Bayat}},
  \bibinfo {author} {\bibfnamefont {S.}~\bibnamefont {Campbell}},\ and\
  \bibinfo {author} {\bibfnamefont {A.~K.}\ \bibnamefont {Mitchell}},\
  }\bibfield  {title} {\bibinfo {title} {Multiparameter critical quantum
  metrology with impurity probes},\ }\href
  {https://doi.org/10.1088/2058-9565/ad438d} {\bibfield  {journal} {\bibinfo
  {journal} {Quantum Sci. Technol.}\ }\textbf {\bibinfo {volume} {9}},\
  \bibinfo {pages} {035033} (\bibinfo {year} {2024}{\natexlab{a}})}\BibitemShut
  {NoStop}%
\bibitem [{\citenamefont {Brunelli}\ \emph {et~al.}(2011)\citenamefont
  {Brunelli}, \citenamefont {Olivares},\ and\ \citenamefont
  {Paris}}]{Brunelli2011PRA}%
  \BibitemOpen
  \bibfield  {author} {\bibinfo {author} {\bibfnamefont {M.}~\bibnamefont
  {Brunelli}}, \bibinfo {author} {\bibfnamefont {S.}~\bibnamefont {Olivares}},\
  and\ \bibinfo {author} {\bibfnamefont {M.~G.~A.}\ \bibnamefont {Paris}},\
  }\bibfield  {title} {\bibinfo {title} {Qubit thermometry for micromechanical
  resonators},\ }\href {https://doi.org/10.1103/PhysRevA.84.032105} {\bibfield
  {journal} {\bibinfo  {journal} {Phys. Rev. A}\ }\textbf {\bibinfo {volume}
  {84}},\ \bibinfo {pages} {032105} (\bibinfo {year} {2011})}\BibitemShut
  {NoStop}%
\bibitem [{\citenamefont {Brunelli}\ \emph {et~al.}(2012)\citenamefont
  {Brunelli}, \citenamefont {Olivares}, \citenamefont {Paternostro},\ and\
  \citenamefont {Paris}}]{Brunelli2012PRA}%
  \BibitemOpen
  \bibfield  {author} {\bibinfo {author} {\bibfnamefont {M.}~\bibnamefont
  {Brunelli}}, \bibinfo {author} {\bibfnamefont {S.}~\bibnamefont {Olivares}},
  \bibinfo {author} {\bibfnamefont {M.}~\bibnamefont {Paternostro}},\ and\
  \bibinfo {author} {\bibfnamefont {M.~G.~A.}\ \bibnamefont {Paris}},\
  }\bibfield  {title} {\bibinfo {title} {Qubit-assisted thermometry of a
  quantum harmonic oscillator},\ }\href
  {https://doi.org/10.1103/PhysRevA.86.012125} {\bibfield  {journal} {\bibinfo
  {journal} {Phys. Rev. A}\ }\textbf {\bibinfo {volume} {86}},\ \bibinfo
  {pages} {012125} (\bibinfo {year} {2012})}\BibitemShut {NoStop}%
\bibitem [{\citenamefont {Cavina}\ \emph {et~al.}(2018)\citenamefont {Cavina},
  \citenamefont {Mancino}, \citenamefont {De~Pasquale}, \citenamefont
  {Gianani}, \citenamefont {Sbroscia}, \citenamefont {Booth}, \citenamefont
  {Roccia}, \citenamefont {Raimondi}, \citenamefont {Giovannetti},\ and\
  \citenamefont {Barbieri}}]{Cavina2018PRA}%
  \BibitemOpen
  \bibfield  {author} {\bibinfo {author} {\bibfnamefont {V.}~\bibnamefont
  {Cavina}}, \bibinfo {author} {\bibfnamefont {L.}~\bibnamefont {Mancino}},
  \bibinfo {author} {\bibfnamefont {A.}~\bibnamefont {De~Pasquale}}, \bibinfo
  {author} {\bibfnamefont {I.}~\bibnamefont {Gianani}}, \bibinfo {author}
  {\bibfnamefont {M.}~\bibnamefont {Sbroscia}}, \bibinfo {author}
  {\bibfnamefont {R.~I.}\ \bibnamefont {Booth}}, \bibinfo {author}
  {\bibfnamefont {E.}~\bibnamefont {Roccia}}, \bibinfo {author} {\bibfnamefont
  {R.}~\bibnamefont {Raimondi}}, \bibinfo {author} {\bibfnamefont
  {V.}~\bibnamefont {Giovannetti}},\ and\ \bibinfo {author} {\bibfnamefont
  {M.}~\bibnamefont {Barbieri}},\ }\bibfield  {title} {\bibinfo {title}
  {Bridging thermodynamics and metrology in nonequilibrium quantum
  thermometry},\ }\href {https://doi.org/10.1103/PhysRevA.98.050101} {\bibfield
   {journal} {\bibinfo  {journal} {Phys. Rev. A}\ }\textbf {\bibinfo {volume}
  {98}},\ \bibinfo {pages} {050101} (\bibinfo {year} {2018})}\BibitemShut
  {NoStop}%
\bibitem [{\citenamefont {Potts}\ \emph {et~al.}(2019)\citenamefont {Potts},
  \citenamefont {Brask},\ and\ \citenamefont
  {Brunner}}]{Potts2019fundamentallimits}%
  \BibitemOpen
  \bibfield  {author} {\bibinfo {author} {\bibfnamefont {P.~P.}\ \bibnamefont
  {Potts}}, \bibinfo {author} {\bibfnamefont {J.~B.}\ \bibnamefont {Brask}},\
  and\ \bibinfo {author} {\bibfnamefont {N.}~\bibnamefont {Brunner}},\
  }\bibfield  {title} {\bibinfo {title} {Fundamental limits on low-temperature
  quantum thermometry with finite resolution},\ }\href
  {https://doi.org/10.22331/q-2019-07-09-161} {\bibfield  {journal} {\bibinfo
  {journal} {{Quantum}}\ }\textbf {\bibinfo {volume} {3}},\ \bibinfo {pages}
  {161} (\bibinfo {year} {2019})}\BibitemShut {NoStop}%
\bibitem [{\citenamefont {Genoni}\ and\ \citenamefont
  {Tufarelli}(2019)}]{Genoni2019JPA}%
  \BibitemOpen
  \bibfield  {author} {\bibinfo {author} {\bibfnamefont {M.~G.}\ \bibnamefont
  {Genoni}}\ and\ \bibinfo {author} {\bibfnamefont {T.}~\bibnamefont
  {Tufarelli}},\ }\bibfield  {title} {\bibinfo {title} {Non-orthogonal bases
  for quantum metrology},\ }\href {https://doi.org/10.1088/1751-8121/ab3fe0}
  {\bibfield  {journal} {\bibinfo  {journal} {J. Phys. A: Math. Theor.}\
  }\textbf {\bibinfo {volume} {52}},\ \bibinfo {pages} {434002} (\bibinfo
  {year} {2019})}\BibitemShut {NoStop}%
\bibitem [{\citenamefont {Montenegro}\ \emph {et~al.}(2020)\citenamefont
  {Montenegro}, \citenamefont {Genoni}, \citenamefont {Bayat},\ and\
  \citenamefont {Paris}}]{Montenegro2020PRR}%
  \BibitemOpen
  \bibfield  {author} {\bibinfo {author} {\bibfnamefont {V.}~\bibnamefont
  {Montenegro}}, \bibinfo {author} {\bibfnamefont {M.~G.}\ \bibnamefont
  {Genoni}}, \bibinfo {author} {\bibfnamefont {A.}~\bibnamefont {Bayat}},\ and\
  \bibinfo {author} {\bibfnamefont {M.~G.~A.}\ \bibnamefont {Paris}},\
  }\bibfield  {title} {\bibinfo {title} {Mechanical oscillator thermometry in
  the nonlinear optomechanical regime},\ }\href
  {https://doi.org/10.1103/PhysRevResearch.2.043338} {\bibfield  {journal}
  {\bibinfo  {journal} {Phys. Rev. Res.}\ }\textbf {\bibinfo {volume} {2}},\
  \bibinfo {pages} {043338} (\bibinfo {year} {2020})}\BibitemShut {NoStop}%
\bibitem [{\citenamefont {Mancino}\ \emph {et~al.}(2020)\citenamefont
  {Mancino}, \citenamefont {Genoni}, \citenamefont {Barbieri},\ and\
  \citenamefont {Paternostro}}]{Mancino2020PRR}%
  \BibitemOpen
  \bibfield  {author} {\bibinfo {author} {\bibfnamefont {L.}~\bibnamefont
  {Mancino}}, \bibinfo {author} {\bibfnamefont {M.~G.}\ \bibnamefont {Genoni}},
  \bibinfo {author} {\bibfnamefont {M.}~\bibnamefont {Barbieri}},\ and\
  \bibinfo {author} {\bibfnamefont {M.}~\bibnamefont {Paternostro}},\
  }\bibfield  {title} {\bibinfo {title} {Nonequilibrium readiness and precision
  of gaussian quantum thermometers},\ }\href
  {https://doi.org/10.1103/PhysRevResearch.2.033498} {\bibfield  {journal}
  {\bibinfo  {journal} {Phys. Rev. Res.}\ }\textbf {\bibinfo {volume} {2}},\
  \bibinfo {pages} {033498} (\bibinfo {year} {2020})}\BibitemShut {NoStop}%
\bibitem [{\citenamefont {Candeloro}\ \emph {et~al.}(2021)\citenamefont
  {Candeloro}, \citenamefont {Razzoli}, \citenamefont {Bordone},\ and\
  \citenamefont {Paris}}]{Candeloro2021PRE}%
  \BibitemOpen
  \bibfield  {author} {\bibinfo {author} {\bibfnamefont {A.}~\bibnamefont
  {Candeloro}}, \bibinfo {author} {\bibfnamefont {L.}~\bibnamefont {Razzoli}},
  \bibinfo {author} {\bibfnamefont {P.}~\bibnamefont {Bordone}},\ and\ \bibinfo
  {author} {\bibfnamefont {M.~G.~A.}\ \bibnamefont {Paris}},\ }\bibfield
  {title} {\bibinfo {title} {Role of topology in determining the precision of a
  finite thermometer},\ }\href {https://doi.org/10.1103/PhysRevE.104.014136}
  {\bibfield  {journal} {\bibinfo  {journal} {Phys. Rev. E}\ }\textbf {\bibinfo
  {volume} {104}},\ \bibinfo {pages} {014136} (\bibinfo {year}
  {2021})}\BibitemShut {NoStop}%
\bibitem [{\citenamefont {Salado-Mejía}\ \emph {et~al.}(2021)\citenamefont
  {Salado-Mejía}, \citenamefont {Román-Ancheyta}, \citenamefont
  {Soto-Eguibar},\ and\ \citenamefont {Moya-Cessa}}]{Salado-Mejía2021QST}%
  \BibitemOpen
  \bibfield  {author} {\bibinfo {author} {\bibfnamefont {M.}~\bibnamefont
  {Salado-Mejía}}, \bibinfo {author} {\bibfnamefont {R.}~\bibnamefont
  {Román-Ancheyta}}, \bibinfo {author} {\bibfnamefont {F.}~\bibnamefont
  {Soto-Eguibar}},\ and\ \bibinfo {author} {\bibfnamefont {H.~M.}\ \bibnamefont
  {Moya-Cessa}},\ }\bibfield  {title} {\bibinfo {title} {Spectroscopy and
  critical quantum thermometry in the ultrastrong coupling regime},\ }\href
  {https://doi.org/10.1088/2058-9565/abdca5} {\bibfield  {journal} {\bibinfo
  {journal} {Quantum Sci. Technol.}\ }\textbf {\bibinfo {volume} {6}},\
  \bibinfo {pages} {025010} (\bibinfo {year} {2021})}\BibitemShut {NoStop}%
\bibitem [{\citenamefont {Brenes}\ and\ \citenamefont
  {Segal}(2023)}]{Brenes2023PRA}%
  \BibitemOpen
  \bibfield  {author} {\bibinfo {author} {\bibfnamefont {M.}~\bibnamefont
  {Brenes}}\ and\ \bibinfo {author} {\bibfnamefont {D.}~\bibnamefont {Segal}},\
  }\bibfield  {title} {\bibinfo {title} {Multispin probes for thermometry in
  the strong-coupling regime},\ }\href
  {https://doi.org/10.1103/PhysRevA.108.032220} {\bibfield  {journal} {\bibinfo
   {journal} {Phys. Rev. A}\ }\textbf {\bibinfo {volume} {108}},\ \bibinfo
  {pages} {032220} (\bibinfo {year} {2023})}\BibitemShut {NoStop}%
\bibitem [{\citenamefont {Sone}\ \emph {et~al.}(2024)\citenamefont {Sone},
  \citenamefont {Soares-Pinto},\ and\ \citenamefont {Deffner}}]{Sone2024QST}%
  \BibitemOpen
  \bibfield  {author} {\bibinfo {author} {\bibfnamefont {A.}~\bibnamefont
  {Sone}}, \bibinfo {author} {\bibfnamefont {D.~O.}\ \bibnamefont
  {Soares-Pinto}},\ and\ \bibinfo {author} {\bibfnamefont {S.}~\bibnamefont
  {Deffner}},\ }\bibfield  {title} {\bibinfo {title} {Conditional quantum
  thermometry—enhancing precision by measuring less},\ }\href
  {https://doi.org/10.1088/2058-9565/ad6736} {\bibfield  {journal} {\bibinfo
  {journal} {Quantum Sci. Technol.}\ }\textbf {\bibinfo {volume} {9}},\
  \bibinfo {pages} {045018} (\bibinfo {year} {2024})}\BibitemShut {NoStop}%
\bibitem [{\citenamefont {Aiache}\ \emph {et~al.}(2024)\citenamefont {Aiache},
  \citenamefont {Seida}, \citenamefont {El~Anouz},\ and\ \citenamefont
  {El~Allati}}]{Aiache2024PRE}%
  \BibitemOpen
  \bibfield  {author} {\bibinfo {author} {\bibfnamefont {Y.}~\bibnamefont
  {Aiache}}, \bibinfo {author} {\bibfnamefont {C.}~\bibnamefont {Seida}},
  \bibinfo {author} {\bibfnamefont {K.}~\bibnamefont {El~Anouz}},\ and\
  \bibinfo {author} {\bibfnamefont {A.}~\bibnamefont {El~Allati}},\ }\bibfield
  {title} {\bibinfo {title} {Non-markovian enhancement of nonequilibrium
  quantum thermometry},\ }\href {https://doi.org/10.1103/PhysRevE.110.024132}
  {\bibfield  {journal} {\bibinfo  {journal} {Phys. Rev. E}\ }\textbf {\bibinfo
  {volume} {110}},\ \bibinfo {pages} {024132} (\bibinfo {year}
  {2024})}\BibitemShut {NoStop}%
\bibitem [{\citenamefont {Liu}\ \emph {et~al.}(2019)\citenamefont {Liu},
  \citenamefont {Yuan}, \citenamefont {Lu},\ and\ \citenamefont
  {Wang}}]{Liu2020JPA}%
  \BibitemOpen
  \bibfield  {author} {\bibinfo {author} {\bibfnamefont {J.}~\bibnamefont
  {Liu}}, \bibinfo {author} {\bibfnamefont {H.}~\bibnamefont {Yuan}}, \bibinfo
  {author} {\bibfnamefont {X.-M.}\ \bibnamefont {Lu}},\ and\ \bibinfo {author}
  {\bibfnamefont {X.}~\bibnamefont {Wang}},\ }\bibfield  {title} {\bibinfo
  {title} {Quantum fisher information matrix and multiparameter estimation},\
  }\href {https://doi.org/10.1088/1751-8121/ab5d4d} {\bibfield  {journal}
  {\bibinfo  {journal} {J. Phys. A: Math. Theor.}\ }\textbf {\bibinfo {volume}
  {53}},\ \bibinfo {pages} {023001} (\bibinfo {year} {2019})}\BibitemShut
  {NoStop}%
\bibitem [{\citenamefont {Mandelstam}\ and\ \citenamefont
  {Tamm}(1945)}]{Mandelstam1945}%
  \BibitemOpen
  \bibfield  {author} {\bibinfo {author} {\bibfnamefont {L.}~\bibnamefont
  {Mandelstam}}\ and\ \bibinfo {author} {\bibfnamefont {I.}~\bibnamefont
  {Tamm}},\ }\bibfield  {title} {\bibinfo {title} {The uncertainty relation
  between energy and time in nonrelativistic quantum mechanics},\ }\href
  {https://link.springer.com/chapter/10.1007/978-3-642-74626-0_8} {\bibfield
  {journal} {\bibinfo  {journal} {J. Phys.}\ }\textbf {\bibinfo {volume} {9}},\
  \bibinfo {pages} {249} (\bibinfo {year} {1945})}\BibitemShut {NoStop}%
\bibitem [{\citenamefont {Uffink}(1993)}]{Uffink1993}%
  \BibitemOpen
  \bibfield  {author} {\bibinfo {author} {\bibfnamefont {J.}~\bibnamefont
  {Uffink}},\ }\bibfield  {title} {\bibinfo {title} {The rate of evolution of a
  quantum state},\ }\href {10.1119/1.17368} {\bibfield  {journal} {\bibinfo
  {journal} {Am. J. Phys.}\ }\textbf {\bibinfo {volume} {61}},\ \bibinfo
  {pages} {935} (\bibinfo {year} {1993})}\BibitemShut {NoStop}%
\bibitem [{\citenamefont {Margolus}\ and\ \citenamefont
  {Levitin}(1998)}]{margolus98}%
  \BibitemOpen
  \bibfield  {author} {\bibinfo {author} {\bibfnamefont {N.}~\bibnamefont
  {Margolus}}\ and\ \bibinfo {author} {\bibfnamefont {L.~B.}\ \bibnamefont
  {Levitin}},\ }\bibfield  {title} {\bibinfo {title} {The maximum speed of
  dynamical evolution},\ }\href {https://doi.org/10.1016/S0167-2789(98)00054-2}
  {\bibfield  {journal} {\bibinfo  {journal} {Physica D}\ }\textbf {\bibinfo
  {volume} {120}},\ \bibinfo {pages} {188} (\bibinfo {year}
  {1998})}\BibitemShut {NoStop}%
\bibitem [{\citenamefont {Levitin}\ and\ \citenamefont
  {Toffoli}(2009)}]{Levitin2009}%
  \BibitemOpen
  \bibfield  {author} {\bibinfo {author} {\bibfnamefont {L.~B.}\ \bibnamefont
  {Levitin}}\ and\ \bibinfo {author} {\bibfnamefont {Y.}~\bibnamefont
  {Toffoli}},\ }\bibfield  {title} {\bibinfo {title} {Fundamental limit on the
  rate of quantum dynamics: {T}he unified bound is tight},\ }\href
  {https://doi.org/10.1103/PhysRevLett.103.160502} {\bibfield  {journal}
  {\bibinfo  {journal} {Phys. Rev. Lett.}\ }\textbf {\bibinfo {volume} {103}},\
  \bibinfo {pages} {160502} (\bibinfo {year} {2009})}\BibitemShut {NoStop}%
\bibitem [{\citenamefont {Fogarty}\ \emph {et~al.}(2020)\citenamefont
  {Fogarty}, \citenamefont {Deffner}, \citenamefont {Busch},\ and\
  \citenamefont {Campbell}}]{Fogarty2020PRL}%
  \BibitemOpen
  \bibfield  {author} {\bibinfo {author} {\bibfnamefont {T.}~\bibnamefont
  {Fogarty}}, \bibinfo {author} {\bibfnamefont {S.}~\bibnamefont {Deffner}},
  \bibinfo {author} {\bibfnamefont {T.}~\bibnamefont {Busch}},\ and\ \bibinfo
  {author} {\bibfnamefont {S.}~\bibnamefont {Campbell}},\ }\bibfield  {title}
  {\bibinfo {title} {Orthogonality catastrophe as a consequence of the quantum
  speed limit},\ }\href {https://doi.org/10.1103/PhysRevLett.124.110601}
  {\bibfield  {journal} {\bibinfo  {journal} {Phys. Rev. Lett.}\ }\textbf
  {\bibinfo {volume} {124}},\ \bibinfo {pages} {110601} (\bibinfo {year}
  {2020})}\BibitemShut {NoStop}%
\bibitem [{\citenamefont {Poggi}\ \emph {et~al.}(2021)\citenamefont {Poggi},
  \citenamefont {Campbell},\ and\ \citenamefont {Deffner}}]{Poggi2021PRQ}%
  \BibitemOpen
  \bibfield  {author} {\bibinfo {author} {\bibfnamefont {P.~M.}\ \bibnamefont
  {Poggi}}, \bibinfo {author} {\bibfnamefont {S.}~\bibnamefont {Campbell}},\
  and\ \bibinfo {author} {\bibfnamefont {S.}~\bibnamefont {Deffner}},\
  }\bibfield  {title} {\bibinfo {title} {Diverging quantum speed limits: A
  herald of classicality},\ }\href
  {https://doi.org/10.1103/PRXQuantum.2.040349} {\bibfield  {journal} {\bibinfo
   {journal} {PRX Quantum}\ }\textbf {\bibinfo {volume} {2}},\ \bibinfo {pages}
  {040349} (\bibinfo {year} {2021})}\BibitemShut {NoStop}%
\bibitem [{\citenamefont {del Campo}(2021)}]{Campo2021PRL}%
  \BibitemOpen
  \bibfield  {author} {\bibinfo {author} {\bibfnamefont {A.}~\bibnamefont {del
  Campo}},\ }\bibfield  {title} {\bibinfo {title} {Probing quantum speed limits
  with ultracold gases},\ }\href
  {https://doi.org/10.1103/PhysRevLett.126.180603} {\bibfield  {journal}
  {\bibinfo  {journal} {Phys. Rev. Lett.}\ }\textbf {\bibinfo {volume} {126}},\
  \bibinfo {pages} {180603} (\bibinfo {year} {2021})}\BibitemShut {NoStop}%
\bibitem [{\citenamefont {Pires}\ \emph {et~al.}(2016)\citenamefont {Pires},
  \citenamefont {Cianciaruso}, \citenamefont {C\'eleri}, \citenamefont
  {Adesso},\ and\ \citenamefont {Soares-Pinto}}]{Pires2016PRX}%
  \BibitemOpen
  \bibfield  {author} {\bibinfo {author} {\bibfnamefont {D.~P.}\ \bibnamefont
  {Pires}}, \bibinfo {author} {\bibfnamefont {M.}~\bibnamefont {Cianciaruso}},
  \bibinfo {author} {\bibfnamefont {L.~C.}\ \bibnamefont {C\'eleri}}, \bibinfo
  {author} {\bibfnamefont {G.}~\bibnamefont {Adesso}},\ and\ \bibinfo {author}
  {\bibfnamefont {D.~O.}\ \bibnamefont {Soares-Pinto}},\ }\bibfield  {title}
  {\bibinfo {title} {Generalized geometric quantum speed limits},\ }\href
  {https://doi.org/10.1103/PhysRevX.6.021031} {\bibfield  {journal} {\bibinfo
  {journal} {Phys. Rev. X}\ }\textbf {\bibinfo {volume} {6}},\ \bibinfo {pages}
  {021031} (\bibinfo {year} {2016})}\BibitemShut {NoStop}%
\bibitem [{\citenamefont {O'Connor}\ \emph {et~al.}(2021)\citenamefont
  {O'Connor}, \citenamefont {Guarnieri},\ and\ \citenamefont
  {Campbell}}]{OConnor2021PRA}%
  \BibitemOpen
  \bibfield  {author} {\bibinfo {author} {\bibfnamefont {E.}~\bibnamefont
  {O'Connor}}, \bibinfo {author} {\bibfnamefont {G.}~\bibnamefont
  {Guarnieri}},\ and\ \bibinfo {author} {\bibfnamefont {S.}~\bibnamefont
  {Campbell}},\ }\bibfield  {title} {\bibinfo {title} {Action quantum speed
  limits},\ }\href {https://doi.org/10.1103/PhysRevA.103.022210} {\bibfield
  {journal} {\bibinfo  {journal} {Phys. Rev. A}\ }\textbf {\bibinfo {volume}
  {103}},\ \bibinfo {pages} {022210} (\bibinfo {year} {2021})}\BibitemShut
  {NoStop}%
\bibitem [{\citenamefont {Deffner}(2017)}]{Deffner2017NJP}%
  \BibitemOpen
  \bibfield  {author} {\bibinfo {author} {\bibfnamefont {S.}~\bibnamefont
  {Deffner}},\ }\bibfield  {title} {\bibinfo {title} {Geometric quantum speed
  limits: a case for {Wigner} phase space},\ }\href
  {https://doi.org/10.1088/1367-2630/aa83dc} {\bibfield  {journal} {\bibinfo
  {journal} {New J. Phys.}\ }\textbf {\bibinfo {volume} {19}},\ \bibinfo
  {pages} {103018} (\bibinfo {year} {2017})}\BibitemShut {NoStop}%
\bibitem [{\citenamefont {Deffner}(2020)}]{Deffner2020PRR}%
  \BibitemOpen
  \bibfield  {author} {\bibinfo {author} {\bibfnamefont {S.}~\bibnamefont
  {Deffner}},\ }\bibfield  {title} {\bibinfo {title} {Quantum speed limits and
  the maximal rate of information production},\ }\href@noop {} {\bibfield
  {journal} {\bibinfo  {journal} {Phys. Rev. Research}\ }\textbf {\bibinfo
  {volume} {2}},\ \bibinfo {pages} {013161} (\bibinfo {year}
  {2020})}\BibitemShut {NoStop}%
\bibitem [{\citenamefont {Aifer}\ and\ \citenamefont
  {Deffner}(2022)}]{Aifer2022NJP}%
  \BibitemOpen
  \bibfield  {author} {\bibinfo {author} {\bibfnamefont {M.}~\bibnamefont
  {Aifer}}\ and\ \bibinfo {author} {\bibfnamefont {S.}~\bibnamefont
  {Deffner}},\ }\bibfield  {title} {\bibinfo {title} {From quantum speed limits
  to energy-efficient quantum gates},\ }\href
  {https://doi.org/10.1088/1367-2630/ac6821} {\bibfield  {journal} {\bibinfo
  {journal} {New J. Phys.}\ }\textbf {\bibinfo {volume} {24}},\ \bibinfo
  {pages} {055002} (\bibinfo {year} {2022})}\BibitemShut {NoStop}%
\bibitem [{\citenamefont {Gross}(1961)}]{Gross1961}%
  \BibitemOpen
  \bibfield  {author} {\bibinfo {author} {\bibfnamefont {E.~P.}\ \bibnamefont
  {Gross}},\ }\bibfield  {title} {\bibinfo {title} {{Structure of a quantized
  vortex in boson systems}},\ }\href
  {https://link.springer.com/article/10.1007/BF02731494} {\bibfield  {journal}
  {\bibinfo  {journal} {Nuovo Cim.}\ }\textbf {\bibinfo {volume} {20}},\
  \bibinfo {pages} {454} (\bibinfo {year} {1961})}\BibitemShut {NoStop}%
\bibitem [{\citenamefont {Pitaevskii}(1961)}]{Pitaevskii1961}%
  \BibitemOpen
  \bibfield  {author} {\bibinfo {author} {\bibfnamefont {L.~P.}\ \bibnamefont
  {Pitaevskii}},\ }\bibfield  {title} {\bibinfo {title} {{Vortex lines in an
  imperfect Bose gas}},\ }\href
  {http://www.jetp.ras.ru/cgi-bin/e/index/e/13/2/p451?a=list} {\bibfield
  {journal} {\bibinfo  {journal} {Sov. J. Exp. Theor. Phys.}\ }\textbf
  {\bibinfo {volume} {13}},\ \bibinfo {pages} {451} (\bibinfo {year}
  {1961})}\BibitemShut {NoStop}%
\bibitem [{\citenamefont {Kolomeisky}\ \emph {et~al.}(2000)\citenamefont
  {Kolomeisky}, \citenamefont {Newman}, \citenamefont {Straley},\ and\
  \citenamefont {Qi}}]{Kolomeisky2000PRL}%
  \BibitemOpen
  \bibfield  {author} {\bibinfo {author} {\bibfnamefont {E.~B.}\ \bibnamefont
  {Kolomeisky}}, \bibinfo {author} {\bibfnamefont {T.~J.}\ \bibnamefont
  {Newman}}, \bibinfo {author} {\bibfnamefont {J.~P.}\ \bibnamefont
  {Straley}},\ and\ \bibinfo {author} {\bibfnamefont {X.}~\bibnamefont {Qi}},\
  }\bibfield  {title} {\bibinfo {title} {Low-dimensional bose liquids: Beyond
  the gross-pitaevskii approximation},\ }\href {10.1103/PhysRevLett.85.1146}
  {\bibfield  {journal} {\bibinfo  {journal} {Phys. Rev. Lett.}\ }\textbf
  {\bibinfo {volume} {85}},\ \bibinfo {pages} {1146} (\bibinfo {year}
  {2000})}\BibitemShut {NoStop}%
\bibitem [{\citenamefont {Barnes}\ and\ \citenamefont
  {Das~Sarma}(2012)}]{Barnes2012PRL}%
  \BibitemOpen
  \bibfield  {author} {\bibinfo {author} {\bibfnamefont {E.}~\bibnamefont
  {Barnes}}\ and\ \bibinfo {author} {\bibfnamefont {S.}~\bibnamefont
  {Das~Sarma}},\ }\bibfield  {title} {\bibinfo {title} {Analytically solvable
  driven time-dependent two-level quantum systems},\ }\href
  {https://doi.org/10.1103/PhysRevLett.109.060401} {\bibfield  {journal}
  {\bibinfo  {journal} {Phys. Rev. Lett.}\ }\textbf {\bibinfo {volume} {109}},\
  \bibinfo {pages} {060401} (\bibinfo {year} {2012})}\BibitemShut {NoStop}%
\bibitem [{\citenamefont {Barnes}(2013)}]{Barnes2013PRA}%
  \BibitemOpen
  \bibfield  {author} {\bibinfo {author} {\bibfnamefont {E.}~\bibnamefont
  {Barnes}},\ }\bibfield  {title} {\bibinfo {title} {Analytically solvable
  two-level quantum systems and landau-zener interferometry},\ }\href
  {https://doi.org/10.1103/PhysRevA.88.013818} {\bibfield  {journal} {\bibinfo
  {journal} {Phys. Rev. A}\ }\textbf {\bibinfo {volume} {88}},\ \bibinfo
  {pages} {013818} (\bibinfo {year} {2013})}\BibitemShut {NoStop}%
\bibitem [{\citenamefont {Mihailescu}\ \emph
  {et~al.}(2024{\natexlab{b}})\citenamefont {Mihailescu}, \citenamefont
  {Campbell},\ and\ \citenamefont {Gietka}}]{mihailescu2024uncertain}%
  \BibitemOpen
  \bibfield  {author} {\bibinfo {author} {\bibfnamefont {G.}~\bibnamefont
  {Mihailescu}}, \bibinfo {author} {\bibfnamefont {S.}~\bibnamefont
  {Campbell}},\ and\ \bibinfo {author} {\bibfnamefont {K.}~\bibnamefont
  {Gietka}},\ }\bibfield  {title} {\bibinfo {title} {Uncertain quantum critical
  metrology: From single to multi parameter sensing},\ }\href@noop {}
  {\bibfield  {journal} {\bibinfo  {journal} {arXiv preprint arXiv:2407.19917}\
  } (\bibinfo {year} {2024}{\natexlab{b}})}\BibitemShut {NoStop}%
\bibitem [{\citenamefont {Childs}\ and\ \citenamefont
  {Young}(2016)}]{Childs2016PRA}%
  \BibitemOpen
  \bibfield  {author} {\bibinfo {author} {\bibfnamefont {A.~M.}\ \bibnamefont
  {Childs}}\ and\ \bibinfo {author} {\bibfnamefont {J.}~\bibnamefont {Young}},\
  }\bibfield  {title} {\bibinfo {title} {Optimal state discrimination and
  unstructured search in nonlinear quantum mechanics},\ }\href
  {https://doi.org/10.1103/PhysRevA.93.022314} {\bibfield  {journal} {\bibinfo
  {journal} {Phys. Rev. A}\ }\textbf {\bibinfo {volume} {93}},\ \bibinfo
  {pages} {022314} (\bibinfo {year} {2016})}\BibitemShut {NoStop}%
\bibitem [{\citenamefont {Rand}(2010)}]{Rand2010}%
  \BibitemOpen
  \bibfield  {author} {\bibinfo {author} {\bibfnamefont {S.}~\bibnamefont
  {Rand}},\ }\href
  {http://books.google.com/books?hl=en{\&}lr={\&}id=dMKc6N9gVs4C{\&}oi=fnd{\&}pg=PR9{\&}dq=Nonlinear+and+Quantum+Optics{\&}ots=uHnGn43Dr{\_}{\&}sig=1tc7diqscyqaccefDaCIV4L2Njk}
  {\emph {\bibinfo {title} {{Nonlinear and Quantum Optics using the density
  matrix}}}}\ (\bibinfo  {publisher} {Oxford University Press},\ \bibinfo
  {year} {2010})\BibitemShut {NoStop}%
\bibitem [{\citenamefont {Ruderman}(2002)}]{Ruderman2002}%
  \BibitemOpen
  \bibfield  {author} {\bibinfo {author} {\bibfnamefont {M.~S.}\ \bibnamefont
  {Ruderman}},\ }\bibfield  {title} {\bibinfo {title} {{Propagation of solitons
  of the Derivative Nonlinear Schr\"odinger equation in a plasma with
  fluctuating density}},\ }\href {https://doi.org/10.1063/1.1482764} {\bibfield
   {journal} {\bibinfo  {journal} {Phys. Plasmas}\ }\textbf {\bibinfo {volume}
  {9}},\ \bibinfo {pages} {2940} (\bibinfo {year} {2002})}\BibitemShut
  {NoStop}%
\bibitem [{\citenamefont {Meyer}\ and\ \citenamefont
  {Wong}(2014)}]{Meyer2014PRA}%
  \BibitemOpen
  \bibfield  {author} {\bibinfo {author} {\bibfnamefont {D.~A.}\ \bibnamefont
  {Meyer}}\ and\ \bibinfo {author} {\bibfnamefont {T.~G.}\ \bibnamefont
  {Wong}},\ }\bibfield  {title} {\bibinfo {title} {Quantum search with general
  nonlinearities},\ }\href {https://doi.org/10.1103/PhysRevA.89.012312}
  {\bibfield  {journal} {\bibinfo  {journal} {Phys. Rev. A}\ }\textbf {\bibinfo
  {volume} {89}},\ \bibinfo {pages} {012312} (\bibinfo {year}
  {2014})}\BibitemShut {NoStop}%
\bibitem [{\citenamefont {Landau}(1932)}]{landau1932theorie}%
  \BibitemOpen
  \bibfield  {author} {\bibinfo {author} {\bibfnamefont {L.}~\bibnamefont
  {Landau}},\ }\bibfield  {title} {\bibinfo {title} {Zur theorie der
  energieubertragung. ii},\ }\href@noop {} {\bibfield  {journal} {\bibinfo
  {journal} {Physikalische Zeitschrift der Sowjetunion}\ }\textbf {\bibinfo
  {volume} {2}},\ \bibinfo {pages} {46} (\bibinfo {year} {1932})}\BibitemShut
  {NoStop}%
\bibitem [{\citenamefont {Zener}(1932)}]{zener1932non}%
  \BibitemOpen
  \bibfield  {author} {\bibinfo {author} {\bibfnamefont {C.}~\bibnamefont
  {Zener}},\ }\bibfield  {title} {\bibinfo {title} {Non-adiabatic crossing of
  energy levels},\ }\href@noop {} {\bibfield  {journal} {\bibinfo  {journal}
  {Proceedings of the Royal Society of London. Series A, Containing Papers of a
  Mathematical and Physical Character}\ }\textbf {\bibinfo {volume} {137}},\
  \bibinfo {pages} {696} (\bibinfo {year} {1932})}\BibitemShut {NoStop}%
\bibitem [{\citenamefont {St{\"u}ckelberg}(1932)}]{stuckelberg1932theorie}%
  \BibitemOpen
  \bibfield  {author} {\bibinfo {author} {\bibfnamefont {E.}~\bibnamefont
  {St{\"u}ckelberg}},\ }\bibfield  {title} {\bibinfo {title} {Theorie der
  unelastischen st{\"o}sse zwischen atomen},\ }\href@noop {} {\bibfield
  {journal} {\bibinfo  {journal} {Helv. Phys. Acta}\ }\textbf {\bibinfo
  {volume} {5}},\ \bibinfo {pages} {369} (\bibinfo {year} {1932})}\BibitemShut
  {NoStop}%
\bibitem [{\citenamefont {Byrnes}\ \emph {et~al.}(2015)\citenamefont {Byrnes},
  \citenamefont {Rosseau}, \citenamefont {Khosla}, \citenamefont {Pyrkov},
  \citenamefont {Thomasen}, \citenamefont {Mukai}, \citenamefont {Koyama},
  \citenamefont {Abdelrahman},\ and\ \citenamefont {Ilo-Okeke}}]{Byrnes2015}%
  \BibitemOpen
  \bibfield  {author} {\bibinfo {author} {\bibfnamefont {T.}~\bibnamefont
  {Byrnes}}, \bibinfo {author} {\bibfnamefont {D.}~\bibnamefont {Rosseau}},
  \bibinfo {author} {\bibfnamefont {M.}~\bibnamefont {Khosla}}, \bibinfo
  {author} {\bibfnamefont {A.}~\bibnamefont {Pyrkov}}, \bibinfo {author}
  {\bibfnamefont {A.}~\bibnamefont {Thomasen}}, \bibinfo {author}
  {\bibfnamefont {T.}~\bibnamefont {Mukai}}, \bibinfo {author} {\bibfnamefont
  {S.}~\bibnamefont {Koyama}}, \bibinfo {author} {\bibfnamefont
  {A.}~\bibnamefont {Abdelrahman}},\ and\ \bibinfo {author} {\bibfnamefont
  {E.}~\bibnamefont {Ilo-Okeke}},\ }\bibfield  {title} {\bibinfo {title}
  {Macroscopic quantum information processing using spin coherent states},\
  }\href {https://doi.org/https://doi.org/10.1016/j.optcom.2014.08.017}
  {\bibfield  {journal} {\bibinfo  {journal} {Optics Communications}\ }\textbf
  {\bibinfo {volume} {337}},\ \bibinfo {pages} {102} (\bibinfo {year}
  {2015})},\ \bibinfo {note} {macroscopic quantumness: theory and applications
  in optical sciences}\BibitemShut {NoStop}%
\bibitem [{\citenamefont {Deffner}\ and\ \citenamefont
  {Lutz}(2013)}]{Deffner2013PRL}%
  \BibitemOpen
  \bibfield  {author} {\bibinfo {author} {\bibfnamefont {S.}~\bibnamefont
  {Deffner}}\ and\ \bibinfo {author} {\bibfnamefont {E.}~\bibnamefont {Lutz}},\
  }\bibfield  {title} {\bibinfo {title} {Quantum speed limit for non-markovian
  dynamics},\ }\href {https://doi.org/10.1103/PhysRevLett.111.010402}
  {\bibfield  {journal} {\bibinfo  {journal} {Phys. Rev. Lett.}\ }\textbf
  {\bibinfo {volume} {111}},\ \bibinfo {pages} {010402} (\bibinfo {year}
  {2013})}\BibitemShut {NoStop}%
\bibitem [{\citenamefont {Cimmarusti}\ \emph {et~al.}(2015)\citenamefont
  {Cimmarusti}, \citenamefont {Yan}, \citenamefont {Patterson}, \citenamefont
  {Corcos}, \citenamefont {Orozco},\ and\ \citenamefont
  {Deffner}}]{Cimmarusti2015PRL}%
  \BibitemOpen
  \bibfield  {author} {\bibinfo {author} {\bibfnamefont {A.~D.}\ \bibnamefont
  {Cimmarusti}}, \bibinfo {author} {\bibfnamefont {Z.}~\bibnamefont {Yan}},
  \bibinfo {author} {\bibfnamefont {B.~D.}\ \bibnamefont {Patterson}}, \bibinfo
  {author} {\bibfnamefont {L.~P.}\ \bibnamefont {Corcos}}, \bibinfo {author}
  {\bibfnamefont {L.~A.}\ \bibnamefont {Orozco}},\ and\ \bibinfo {author}
  {\bibfnamefont {S.}~\bibnamefont {Deffner}},\ }\bibfield  {title} {\bibinfo
  {title} {Environment-assisted speed-up of the field evolution in cavity
  quantum electrodynamics},\ }\href
  {https://doi.org/10.1103/PhysRevLett.114.233602} {\bibfield  {journal}
  {\bibinfo  {journal} {Phys. Rev. Lett.}\ }\textbf {\bibinfo {volume} {114}},\
  \bibinfo {pages} {233602} (\bibinfo {year} {2015})}\BibitemShut {NoStop}%
\end{thebibliography}%

\end{document}